\documentclass[12pt,preprintnumbers,superscriptaddress,amsmath,amssymb,nofootinbib]{revtex4-1}

\usepackage{graphicx}
\usepackage{dcolumn}
\usepackage{bm}
\usepackage{amssymb}
\usepackage{amsmath}
\usepackage{epsfig}    
\usepackage{color}
\usepackage{slashed}
\usepackage{hyperref}
\usepackage{float}

\begin{document} 
	
\title{CP-Violating 2HDMs Emerging from 3-3-1 Models}
\preprint{OU-HET-1129}
\author{Zhiyi Fan}
\email{fzy1136387253@gmail.com}
\author{Kei Yagyu}
\email{yagyu@het.phys.sci.osaka-u.ac.jp}
\affiliation{Department of Physics, Osaka University, Toyonaka, Osaka 560-0043, Japan}
	
\begin{abstract}

We investigate CP violating 2 Higgs doublet models as an effective theory emerging from models with an $SU(3)_L \otimes U(1)_X$ gauge symmetry. 
Because of the extension of the electroweak symmetry, a characteristic structure of Yukawa interactions appears with new CP violating phases at the electroweak scale. 
In this scenario, the charged Higgs boson loop provides dominant and sizable contributions to the neutron electric dipole moment (nEDM) at one-loop level. 
We find that the prediction of the nEDM can be slightly smaller than the current upper limit, ${\cal O}(10^{-26})\,e\,\text{cm}$, with the mass of the charged Higgs boson and new CP violating phases to be a few hundred GeV and $\mathcal{O}(1)$, respectively, under the constraints from the meson mixings, the $B \to X_s \gamma$ decay and current LHC data. 

\end{abstract}
	\maketitle

\section{Introduction}

The three-generation structure of matter fields assumed in the standard model (SM) has been established by the discovery of top quarks at Tevatron~\cite{CDF:1995wbb,D0:1995jca}. 
In addition, its consistency has precisely been tested by various experiments such as $B$ factories~\cite{BaBar:2014omp} for the quark sector and LEP for the lepton sector~\cite{ALEPH:2005ab}. 
Furthermore, CP violation (CPV) is naturally induced from the phase of the Cabibbo-Kobayashi-Maskawa (CKM) matrix, which is required to generate non-zero baryon asymmetry of the Universe.  
In spite of such phenomenological successes, there is no theoretical basis for the existence of the three generations in the SM, and it has been well known that 
the amount of CPV from the CKM matrix is not sufficient to explain the observed baryon asymmetry~\cite{Huet:1994jb}. 
These problems are expected to be solved in new physics scenarios beyond the SM. 

Models with an extended electroweak (EW) gauge symmetry $SU(3)_C\otimes SU(3)_L \otimes U(1)_X$, the so-called 3-3-1 models~\cite{Singer:1980sw,Valle:1983dk,Frampton:1992wt}, 
can simultaneously explain the origin of three generations and additional sources of CPV. 
The former is deduced from the condition for the gauge anomaly cancellation, by which 
the number of generations has to be proportional to the color degrees of freedom~\cite{Frampton:1992wt}. 
The extension of the EW symmetry naturally requires an extension of the Higgs sector, minimally containing three $SU(3)_L$ triplet Higgs fields,  
in order to realize the spontaneous symmetry breaking $SU(3)_L \otimes U(1)_X \to U(1)_{\rm em}$ and make all charged fermions massive. 
Such an extended Higgs sector leads to a richer structure of Yukawa interactions and the Higgs potential with additional sources of CPV. 

In this paper, we discuss an effective theory described by the $SU(2)_L \otimes U(1)_Y$ symmetry which is deduced from various 3-3-1 models with the minimal Higgs sector. 
All the non-SM fermions and gauge bosons are decoupled from the theory by taking the vacuum expectation value (VEV) of $SU(3)_L \otimes U(1)_X \to SU(2)_L \otimes U(1)_Y$ to be infinity. 
Even in this limit, the Higgs sector can be non-minimal including two $SU(2)_L$ doublet fields, namely the effective theory corresponds to a two Higgs doublet model (2HDM)~\cite{Okada:2016whh}. 
This 2HDM is categorized into that with a discrete $Z_2$ symmetry whose charge for SM quarks are flavor dependent, not the flavor universal one known as the Type-I, Type-II, Type-X and Type-Y 2HDMs~\cite{Grossman:1994jb,Barger:2009me,Aoki:2009ha}. 
Because of the flavor dependent structure, Higgs boson couplings with quarks include new sources of flavor violation and CPV phases other than those from the CKM matrix, 
which can give sizable contributions to $B$ physics observables and electric dipole moments (EDMs). 
We show that the one-loop diagram with charged Higgs boson and top quark loops can give a significant contribution to the neutron EDM (nEDM), 
by which a portion of the parameter space is excluded by the current measurement or explored by future experiments such as the n2EDM~\cite{n2EDM:2021yah}
\footnote{CPV effects on EDMs in 3-3-1 models have also been discussed in Refs.~\cite{DeConto:2014fza,DeConto:2016osh}, 
in which effects of extra fermions and gauge bosons are kept by taking their masses to be of order 1 TeV. }. 
This result is quite different from that in the four types of 2HDMs mentioned above~\cite{Abe:2013qla,Cheung:2020ugr,Altmannshofer:2020shb} 
and the so-called aligned 2HDMs~\cite{Pich:2009sp,Jung:2013hka,Kanemura:2020ibp}, where 
one-loop contributions are negligibly smaller than the two-loop Barr-Zee type contributions~\cite{Barr:1990vd} due to chiral suppressions. 
We also find that the new phases and the mass of the charged Higgs boson can be of order one and a few hundred GeV, respectively, under the constraints from the meson mixings 
($B^0$-$\bar{B}^0$, $D^0$-$\bar{D}^0$ and $K^0$-$\bar{K}^0$), the $B \to X_s \gamma$ decay, EDMs and current LHC data. 

This paper is organized as follows. 
In Sec.~\ref{sec:models}, we classify various 3-3-1 models, and show that their effective theory can be a 2HDM. 
In Sec.~\ref{sec:flavor}, we discuss flavor constraints such as the meson mixings, the $B \to X_s \gamma$ decay and EDMs. 
The numerical evaluations for the nEDM are given in Sec.~\ref{sec:num} under the constraint from the flavor observables. 
Conclusions are given in Sec.~\ref{sec:conc}. 
In Appendix~\ref{sec:app1}, we present formulae for the masses of Higgs bosons and those at the large VEV limit.  
In Appendix~\ref{sec:app2}, details of the other 3-3-1 models are presented. 
In Appendix~\ref{sec:decay-rate}, expressions for decay rates of the additional Higgs bosons are explicitly shown. 

\section{Models \label{sec:models}}

In this section, we first classify 3-3-1 models by the difference of the definition of the electric charge $Q$. 
We then show that their effective theory corresponds to a 2HDM. 

\subsection{Classification of 3-3-1 models}

Without loss of generality, $Q$ is defined as 
\begin{align}
Q \equiv T_3 + Y,~~\text{with}~~ Y \equiv \zeta T_8 + X, \label{eq:em}
\end{align}
where $T_3$ and $T_8$ are the diagonal Gell-Mann matrices:
\begin{align}
T_3 =\frac{1}{2} \text{diag}(1,-1,0),\quad T_8 = \frac{1}{2\sqrt{3}}\text{diag}(1,1,-2).  
\end{align}
As shown in Eq.~(\ref{eq:em}), the hypercharge $Y$ is given as the linear combination of $T_8$ and $X$ with $X$ being the $U(1)_X$ charge. 
In order to make $Y$ a rational number, the $\zeta$ parameter should be proportional to $1/\sqrt{3}$.  
We thus express $\zeta = n/\sqrt{3}$ with $n$ being an integer. 
The electric charge for $SU(3)_L$ triplets $\bm 3$ is then expressed as 
\begin{align}
Q(\bm 3) = \begin{pmatrix}
\frac{1}{2} + \frac{n}{6} + X \\
-\frac{1}{2} + \frac{n}{6} + X \\
-\frac{n}{3} + X
\end{pmatrix}. \label{eq:q3}
\end{align}
The electric charge for the third component field $Q_3$ is expressed by that of the first component $Q_1$ as $Q_3 = Q_1 -(n + 1)/2$, so that 
$n$ should be an odd number to make $Q_3$ integer, by which we can avoid the appearance of exotic lepton fields with $Q$ being half integer. 

The magnitude of $n$ is constrained by taking into account the relation among gauge coupling constants~\cite{Fonseca:2016xsy}. 
Suppose that the $SU(3)_L$ symmetry is spontaneously broken down to $SU(2)_L\otimes U(1)_Y$ by a VEV of a triplet Higgs field $\Phi_3$ with the $X$ charge of $n/3$, 
i.e., $\langle \Phi_3 \rangle = (0,0,v_3/\sqrt{2})^T$\footnote{For $n=\pm 1$, 
the first or second component of $\Phi_3$ is electrically neutral, so that it can generally develop a non-zero VEV. }. 
After the breaking of $SU(3)_L \to SU(2)_L\otimes U(1)_Y$, the 8th component of the $SU(3)_L$ gauge boson $A_8^\mu$ is mixed with the $U(1)_X$ gauge boson $X^\mu$ as 
\begin{align}
\begin{pmatrix}
A_\mu^8 \\
X_\mu
\end{pmatrix}
= 
\begin{pmatrix}
\cos\theta_{331} &  \sin\theta_{331} \\
-\sin\theta_{331} & \cos\theta_{331}
\end{pmatrix}
\begin{pmatrix}
X_\mu' \\
B_\mu
\end{pmatrix}, 
\end{align}
where $B_\mu$ ($X_\mu'$) is the hypercharge (an additional neutral massive) gauge boson. The mixing angle $\theta_{331}$ is given by 
\begin{align}
\sin\theta_{331} = \frac{\zeta g_X}{\sqrt{g^2 + \zeta^2g_X^2}},\quad 
\cos\theta_{331} = \frac{g}{\sqrt{g^2 + \zeta^2g_X^2}}.  \label{eq:333}
\end{align}
The hypercharge coupling $g_Y^{}$ is given to be $g_Y = g\sin\theta_{331} = g_X\cos\theta_{331}$. Using Eq.~(\ref{eq:333}), this can also be expressed as 
\begin{align}
\frac{1}{g_Y^2} =  \frac{\zeta^2}{ g^2} + \frac{1}{g_X^2}. 
\end{align}
This relation tells us the upper limit on $|\zeta|$ which corresponds to the case with $g_X \to \infty$ as 
\begin{align}
\zeta^2 < \frac{g^2}{g^2_Y} = \frac{1}{\tan^2\theta_W} \simeq (1.87)^2. 
\end{align}
Therefore, possible choices for $n$ are $n = \pm 1$ and $\pm 3$, i.e., $\zeta = \pm 1/\sqrt{3}$ and  $\pm \sqrt{3}$. 

3-3-1 models can also be classified into two categories depending on the way how three generations of left-handed quarks $Q_L$ are embedded into the $SU(3)_L$ representation. 
In the first category, denoting Class-I, first two generations $Q_L^{1,2}$ are assigned to be $SU(3)_L$ anti-triplet, while the third generation $Q_L^3$ and three left-handed leptons $L_L^{1,2,3}$ are assigned to be triplet. 
The gauge anomaly for $[SU(3)_L]^3$ is then cancelled between six triplets from $Q_L^3$ and $L_L^{1,2,3}$ and six anti-triplets from $Q_L^{1,2}$. 
In Class-I, we can further decompose models into those with $\zeta=-\sqrt{3}$~\cite{Frampton:1992ui,Pisano_1992}, 
$\zeta=-1/\sqrt{3}$~\cite{Singer:1980uv}, $\zeta=1/\sqrt{3}$~\cite{Pleitez_1996,Ozer:1996fb} and $\zeta=\sqrt{3}$~\cite{Fonseca:2016xsy}.
In another category, denoting Class-II, all three generations of $Q_L$ are embedded into anti-triplets, while the 
lepton sector has to be extended to include nine generations of left-handed lepton triplets, the so-called $E_6$ inspired model~\cite{Sanchez:2001ua}, or 
to include one sextet (contributing to $[SU(3)_L]^3$ as much as the seven triplets) and two triplets, the so-called flipped model~\cite{Fonseca_2016}. 
We note that in Class-II, the $\zeta$ parameter is uniquely determined to be $-1/\sqrt{3}$ for the $E_6$ inspired model and $1/\sqrt{3}$ for the flipped model. 

In the following, we mainly focus on Class-I models with $\zeta=-1/\sqrt{3}$ and its effective theory which is obtained by taking a large limit of the $SU(3)_L$ breaking VEV.  
We then comment on models with the other $\zeta$ values and Class-II models. 

\subsection{Model with $\zeta=-1/\sqrt{3}$}

\begin{table}[t]
	\renewcommand\arraystretch{0.8}
	\begin{center}
		\begin{tabular}{ccccc}
			\hline
			Fields&~~$SU(3)_C\otimes SU(3)_L \otimes U(1)_X\otimes U(1)'$~~&~~$Z_2^{\rm rem}$~~&Components& \\
			\hline
			$Q^a_L$&$\left(\bm{3},\overline{\bm{3}},0,0\right)$& $(+,+,-)$ &$\left((q_L^a)^T,D^a_L\right)^T$,~ $q_L^a\equiv\left(d^a_L,-u^a_L\right)^T$&\\
			$Q^3_L$&$\left(\bm{3},\bm{3},+1/3,0\right)$&$(+,+,-)$&$\left((q_L^3)^T,U_L\right)^T,~q_L^3\equiv\left(t_L,b_L\right)^T$&\\
			$u^i_R$&$\left(\bm{3},\bm{1},+2/3,q\right)$&$+$&$u^i_R$&\\
			$d^i_R$&$\left(\bm{3},\bm{1},-1/3,-q\right)$&$+$&$d^i_R$&\\
			$U_R$&$\left(\bm{3},\bm{1},+2/3,2q\right)$&$-$&$U_R$&\\
			$D^a_R$&$\left(\bm{3},\bm{1},-1/3,-2q\right)$&$-$&$D^a_R$&\\
			$L^i_L$&$\left(\bm{1},\bm{3},-1/3,0\right)$&$(+,+,-)$&$\left((\ell_L^i)^T, N_R^{ic}\right)^T,~\ell_L^i\equiv\left(\nu^i_L,e^i_L\right)^T$&\\
			$e^i_R$&$\left(\bm{1},\bm{1},-1,-q\right)$&$+$&$e^i_R$&\\
			$N^i_R$&$\left(\bm{1},\bm{1},0,+2q\right)$&$-$&$N^i_R$&\\\hline	
                        $\Phi_1$&$\left(\bm{1},\bm{3},+2/3,q\right)$& $(+,+,-)$&$\left(\phi_1^T,\eta_1^+\right)^T,~\phi_1\equiv\left(\phi_1^+,\phi_1^0\right)^T$&\\
			$\Phi_2$&$\left(\bm{1},\bm{3},-1/3,-q\right)$&$(+,+,-)$&$\left(\phi_2^T,\eta_2^0 \right),~\phi_2\equiv\left(\phi_2^0,\phi_2^-\right)^T$&\\
	                $\Phi_3$&$\left(\bm{1},\bm{3},-1/3,-2q\right)$&$(-,-,+)$&$\left(\eta_3^T,\phi_3^0\right)^T,~\eta_3\equiv\left(\eta_3^0,\eta_3^-\right)^T$&\\	\hline
		\end{tabular}
		\caption{Particle content of the model with $\zeta = -1/\sqrt{3}$, where $U(1)'$ is a softly-broken global symmetry and  
                  $Z_2^{\rm rem}$ is a remnant symmetry after the spontaneous breaking of $SU(3)_L \otimes U(1)_X$.  
                  The charge of $Z_2^{\rm rem}$ can be defined as $(-1)^{|Q'/q| +  2\sqrt{3}T_8 +2s}$ with $s$ being the spin and $Q'$ being $U(1)'$ charge.
                  Flavor indices $i$ and $a$ run over 1-3 and 1-2, respectively. 
                  In the last column, the component fields of $SU(3)_L$ triplets are shown, where the first two components are written as the doublet form. 
                  }
		\label{table1}
	\end{center}
\end{table}

In order to discuss the effective theory of 3-3-1 models, we consider the model with $\zeta = -1/\sqrt{3}$ in Class-I as a representative one. 
The particle content of this model is given in Table~\ref{table1}. 
In addition to the gauge symmetry, we introduce a global $U(1)'$ symmetry to avoid mixings between SM fermions and extra fermions at tree level. 
The scalar sector is composed of three $SU(3)_L$ triplets, which corresponds to the minimal form to break $SU(3)_L \otimes U(1)_X$ into $U(1)_{\rm em}$
and to give all the masses of fermions except for active neutrinos $\nu_L^i$. 

The Higgs potential is generally written under the $SU(3)_L\otimes U(1)_X\otimes U(1)'$ symmetry as 
\begin{align}
V &= \sum_{i = 1, 3}m_i^2|\Phi_i|^2 + \left(m_{23}^2\Phi_2^\dagger \Phi_3 + \mu \epsilon_{\alpha\beta\gamma}\Phi_1^\alpha\Phi_2^\beta\Phi_3^\gamma + \text{h.c.}\right) \notag\\
&  + \sum_{i = 1, 3} \frac{\lambda_i}{2}|\Phi_i|^4 +  \sum_{i,j=1,3}^{j>i}\lambda_{ij}|\Phi_i|^2|\Phi_j|^2 +  \sum_{i,j=1,3}^{j>i}\rho_{ij}|\Phi_i^\dagger \Phi_j|^2, \label{eq:pot}
\end{align}
where $m_{23}^2$ and $\mu$ terms softly break the $U(1)'$ symmetry, and the subscripts $\alpha,\beta,\gamma$ $(=1,2,3)$ denote the index for the fundamental representation of $SU(3)_L$. 
By rephasing the scalar fields, all the parameters in the potential are taken to be real without loss of generality. 
We take $m_{23}^2 = 0$ such that the configuration of the VEVs $\langle \phi^0_i\rangle = v_i/\sqrt{2} \neq 0$ and  $\langle \eta^0_j\rangle =  0$ 
is realized from the tadpole conditions, see Appendix~\ref{sec:app1}. 
In this case, the symmetry breaking is realized by the following two steps: 
\begin{align}
SU(3)_L\otimes U(1)_X \xrightarrow[v_3]{} SU(2)_L\otimes U(1)_Y\xrightarrow[v]{} U(1)_{\rm em}, 
\end{align}
where $v \equiv \sqrt{v_1^2 + v_2^2} = (\sqrt{2}G_F)^{-1/2}$ with $G_F$ being the Fermi constant. 
The ratio of these VEVs is parameterized as $\tan\beta\equiv v_2/v_1$. 
In this configuration, a remnant unbroken $Z_2^{\rm rem}$ symmetry appears whose charges are given in Table~\ref{table1}, by which 
the component fields $\phi$ and $\eta$ do not mix with each other. 
In addition, the lightest neutral $Z_2^{\rm rem}$-odd particle can be a candidate of dark matter. 

There are 18 real scalar fields in the three triplets $\Phi_i$, but 8 of 18 fields are absorbed into longitudinal components of the massive gauge bosons. 
Thus, 10 real scalar fields remain as the physical degrees of freedom. 
These can be classified into two pairs of singly-charged Higgs bosons ($H^\pm$ and $\eta^\pm$), two ``CP-odd'' Higgs bosons ($A$, $\eta_I^{}$) and 
four ``CP-even'' Higgs bosons ($h$, $H$, $H_S$, $\eta_R^{}$) with $h$ being identified with the SM-like Higgs boson. \footnote{Here, ``CP-even'' and ``CP-odd'' indicate the definite CP-states
when CPV phases in the Yukawa interaction are taken to be zero. In reality, we cannot define such a CP-state, because the Higgs potential receives CPV effects from Yukawa interactions via radiative corrections.  }
See Appendix~\ref{sec:app1} for expressions of the scalar boson masses and the relation between the mass eigenbases and original ones shown in Table~\ref{table1}. 

All the $Z_2^{\rm rem}$-odd scalar fields and the $H_S$ state are decoupled by taking the $SU(3)_L$ breaking VEV $v_3$ to be infinity. 
We note that extra five gauge bosons (a pair of charged, a complex and a real neutral gauge bosons) are also decoupled by taking this limit, see e.g., Ref.~\cite{Okada:2016whh}. 
On the other hand, the remaining five Higgs bosons, i.e., $H^\pm$, $A$, $H$ and $h$ can be of order the EW scale as long as the 
$M^2 \equiv \mu v_3 /(\sqrt{2}\cos\beta \sin\beta)$ parameter is kept to be (EW scale)$^2$.  
Thus, the case with $M \sim {\cal O}(v)$ the Higgs sector of this model is effectively regarded as a 2HDM. 
The Higgs potential of such an effective 2HDM can be written as  
\begin{align}
V_{\rm eff} &= \mu_1^2|\phi_1|^2+\mu_2^2|\phi_2|^2+\mu_{12}^2[\phi_1^T (i\tau_2)\phi_2 + \text{h.c.}]  \notag\\
&  + \frac{\lambda_1}{2}|\phi_1|^4 + \frac{\lambda_2}{2}|\phi_2|^4 + \lambda_{12}|\phi_1|^2|\phi_2|^2 + \rho_{12}|\phi_1^\dagger \phi_2|^2, \label{eq:pot-2hdm}
\end{align}
where $\phi_{1,2}$ are the $SU(2)_L$ doublet Higgs fields defined in Table~\ref{table1}. The $\mu_{12}^2$ parameter is real, since the original potential does not contain physical CPV phases. 
One can check that the same mass formulae in the limit $v_3 \to \infty$ given in Appendix~\ref{sec:app1} can be reproduced by starting from the above potential with the identification $M^2 = \mu_{12}^2/(s_\beta c_\beta)$. 

The Yukawa interaction is given by 
\begin{align}
\mathcal{L}_Y =
&-(Y_e)_{ij}\overline{L_L^i}\, \Phi_1\, e_R^j -(Y_N)_{ij}\overline{L_L^i}\, \Phi_3\, N_R^j  + \text{h.c.} \notag\\
&+ (Y_u^1)_{ai}\overline{Q_L^a}\, \Phi_1^*\, u_R^i 
- (Y_u^2)_i\overline{Q_L^3}\, \Phi_2\, u_R^i
- Y_U\overline{Q_L^3}\, \Phi_3\, U_R + \text{h.c.} \notag\\
&- (Y_d^1)_i\overline{Q_L^3}\,\Phi_1\, d_R^i 
- (Y_d^2)_{ai}\overline{Q_L^a}\, \Phi_2^*\, d_R^i
- (Y_D)_{ab}\overline{Q_L^a} \,\Phi_3^*\, D_R^b + \text{h.c.}
\end{align}
Due to the $U(1)'$ symmetry, mixings between SM fermions and extra fermions ($U$ and $D^a$) do not appear. 
In addition, the new lepton Yukawa term $\epsilon_{\alpha\beta\gamma} (Y_L)_{ij} (\overline{L_L^{ic}})^\alpha (L_L^{j})^\beta (\Phi_1)^\gamma$ is forbidden. 
We see that the extra fermion masses are given by the large VEV $v_3$, while 
those of the SM fermions are given by the smaller VEVs $v_1$ and $v_2$. 

At the large $v_3$ limit, we obtain the low energy effective Lagrangian as follows: 
\begin{align}
 {\cal L}_{\rm eff} &= -(Y_e)_{ij} \ell_L^i \phi_1 e_R^j + \text{h.c.}\notag\\
 & +(Y_{u}^1)_{ai}\bar{q}_L^a \phi_1^*u_R^i - (Y_{u}^2)_i\bar{q}_L^3 \phi_2 u_R^i
  - (Y_{d}^1)_i\bar{q}_L^3 \phi_1 d_R^i - (Y_{d}^2)_{ai} \bar{q}_L^a \phi_2^* d_R^i + \text{h.c.},  \label{eq:eff}
\end{align}
where $\ell_L^i$, $q_L^a$, $q_L^3$, $\phi_1$ and $\phi_2$ are the $SU(2)_L$ doublet fields defined in Table~\ref{table1}. 
The Yukawa interaction given in Eq.~(\ref{eq:eff}) is invariant under a flavor dependent $Z_2$ transformation, denoting $Z_2^{F}$, whose charge can be defined by $(-1)^{3X}$, by which 
$\{\phi_2,q_L^3,\ell_L,d_R,e_R\}$ are odd under $Z_2^F$, while the other SM fermions and $\phi_1$ are even. 
The $Z_2^F$ symmetry is softly-broken by the $\mu_{12}^2$ term in the potential, see Eq.~(\ref{eq:pot-2hdm}). 
Due to the $Z_2^F$ symmetry, a characteristic flavor dependence in the quark Yukawa interactions appears, which is not seen in the 2HDMs with a usual (flavor independent) $Z_2$ symmetry.
From Eq.~(\ref{eq:eff}), we see that the mass matrices for quarks are composed of two Yukawa matrices as 
\begin{align}
M_u = \frac{1}{\sqrt{2}}
\begin{pmatrix}
v_1(Y_u^1)_{11} & v_1(Y_u^1)_{12} & v_1(Y_u^1)_{13} \\
v_1(Y_u^1)_{21} & v_1(Y_u^1)_{22} & v_1(Y_u^1)_{23}\\
v_2(Y_u^2)_{1} & v_2(Y_u^2)_{2} & v_2(Y_u^2)_{3} \\
\end{pmatrix},\quad 
M_d =
\frac{1}{\sqrt{2}}\begin{pmatrix}
v_2(Y_{d}^2)_{11} & v_2(Y_{d}^2)_{12} & v_2(Y_{d}^2)_{13} \\
v_2(Y_{d}^2)_{21} & v_2(Y_{d}^2)_{22} & v_2(Y_{d}^2)_{23} \\
v_1(Y_{d}^2)_{1} &  v_1(Y_{d}^2)_{2} & v_1(Y_{d}^2)_{3}   \\
\end{pmatrix}. \notag
\end{align}
Therefore, after diagonalizing these mass matrices, flavor dependent structures of the quark Yukawa interactions appear. 
On the other hand, the mass matrix for charged leptons is given by the single Yukawa matrix $Y_e$ as in the SM. 
Performing the following bi-unitary transformations: 
\begin{align}
f_L^{} = V_f f_L',\quad f_R^{} = U_f f_R',~~(f = u,d,e), 
\end{align}
we can diagonalize these mass matrices. 
We then can extract the interaction terms in the mass eigenbases for fermions as follows: 
\begin{align}
{\cal L}_{\rm eff}&= -\frac{1}{v}\bar{e}'M_e^{\rm diag}[(s_{\beta-\alpha} - t_\beta c_{\beta-\alpha})h + (c_{\beta-\alpha} + t_\beta s_{\beta-\alpha})H + i t_\beta \gamma_5A]  e' 
 \notag\\
& -\frac{1}{v}\sum_{\varphi=h,H,A}\sum_{q=u,d}p_\varphi^q\bar{q}'\Gamma_\varphi^q  M_q^{\rm diag} P_R q'\varphi+ \text{h.c.} \notag\\
& -\frac{\sqrt{2}}{v}\left[\bar{e}'M_e^{\rm diag}t_\beta P_L \nu' + \bar{d}'\left( M_L P_L + M_R P_R \right) u'\right]H^- + \text{h.c.},  \label{eq:aa}
\end{align}
where $c_X =\cos X$, $s_X = \sin X$, $t_X = \tan X$, $ p_h^q = p_H^q = 1,~p_A^q = 2iI_q$ with $I_q = 1/2 (-1/2)$ for $q=u(d)$ and 
\begin{align}
M_L = -M_d^{\rm diag} \Gamma_A^d V_{\rm CKM}^\dagger, \quad M_R = V_{\rm CKM}^\dagger \Gamma_A^u M_u^{\rm diag}. \label{eq:mlmr}
\end{align}
In Eq.~(\ref{eq:aa}), $P_{L,R}$ are the projection operator, and $V_{\rm CKM}\equiv V_u^\dagger V_d $ is the CKM matrix. 
The interaction matrices are given by 
\begin{align}
\begin{split}
\Gamma_A^u &= V_u^\dagger \text{diag}(-t_\beta,-t_\beta,t_\beta^{-1})V_u , \quad \Gamma_A^d = V_d^\dagger \text{diag}(t_\beta^{-1},t_\beta^{-1},-t_\beta)V_d,  \\
\Gamma_H^u &= V_u^\dagger \text{diag}\left(c_{\beta-\alpha} + t_\beta s_{\beta-\alpha},c_{\beta-\alpha} + t_\beta s_{\beta-\alpha},c_{\beta-\alpha} - t_\beta^{-1} s_{\beta-\alpha}\right)V_u ,\\
\Gamma_H^d &= V_d^\dagger \text{diag}\left(c_{\beta-\alpha} - t_\beta^{-1} s_{\beta-\alpha},c_{\beta-\alpha} - t_\beta^{-1} s_{\beta-\alpha},c_{\beta-\alpha} + t_\beta s_{\beta-\alpha}\right)V_d , \\
\Gamma_h^u &= V_u^\dagger \text{diag}\left(s_{\beta-\alpha} - t_\beta c_{\beta-\alpha},s_{\beta-\alpha} - t_\beta c_{\beta-\alpha},s_{\beta-\alpha} + t_\beta^{-1} c_{\beta-\alpha}\right)V_u , \\
\Gamma_h^d &= V_d^\dagger \text{diag}\left(s_{\beta-\alpha} + t_\beta^{-1} c_{\beta-\alpha},s_{\beta-\alpha} + t_\beta^{-1} c_{\beta-\alpha},s_{\beta-\alpha} - t_\beta c_{\beta-\alpha}\right)V_d. 
\end{split} \label{eq:gamma0}
\end{align}
All these $\Gamma_\varphi^q$ matrices are hermitian which turn out to be important for the discussion of the EDMs in the next section. 
In the so-called alignment limit, i.e., $s_{\beta-\alpha} \to 1$, these matrices take the simple form as
\begin{align}
\Gamma_H^q &= -\Gamma_A^q,\quad \Gamma_h^q = 1~~(q = u,d). 
\end{align}
Thus, the Yukawa couplings for the 125 GeV Higgs boson $h$ are the same as those of the SM values at tree level. 
On the other hand, the Yukawa couplings for extra Higgs bosons are non-diagonal form, because of the flavor dependent structure of the matrices $\text{diag}(-t_\beta,-t_\beta,t_\beta^{-1})$
and $\text{diag}(t_\beta^{-1},t_\beta^{-1},-t_\beta)$ as aforementioned. 
We note that in the 2HDMs with a flavor independent $Z_2$ symmetry, the diagonal matrices, e.g., $\text{diag}(-t_\beta,-t_\beta,t_\beta^{-1})$ are replaced by the $3\times 3$ identity matrix $I_3$ times a constant, 
so that the above $\Gamma_\varphi^q$ matrices are also proportional to $I_3$ due to the unitarity of the $V_q$ matrices. 

In order to parameterize the $\Gamma_\varphi^q$ matrices, we express the unitary matrix $V_u$ as follows: 
\begin{align}
V_u = \text{diag}(e^{i\delta_1},e^{i\delta_2},e^{i\delta_3})R_{12}(\phi)R_{13}(\theta)R_{23}(\psi)\text{diag}(e^{i\delta_4},e^{i\delta_5},e^{i\delta_6}), 
\end{align}
where 
\begin{align}
R_{12}(X) = \begin{pmatrix}
c_X^{} & -s_X^{} & 0 \\
s_X^{} & c_X^{} & 0 \\
0 & 0 & 1
\end{pmatrix},~
R_{13}(X) = \begin{pmatrix}
c_X^{} &0& -s_X^{}  \\
0 & 1 & 0 \\
s_X^{} &0& c_X^{} 
\end{pmatrix},~
R_{23}(X) = \begin{pmatrix}
1&0&0\\
0 & c_X^{} & -s_X^{}  \\
0 & s_X^{} & c_X^{}
\end{pmatrix}. 
\end{align}
The unitary matrix for down-type quarks $V_d$ is then obtained from $V_d = V_uV_{\rm CKM}$. 
In this parameterization, the $\Gamma_\varphi^{q}$ matrices do not depend on the $(\delta_1,\delta_2,\delta_3,\phi)$ parameters. 

\begin{table}[t]
	\renewcommand\arraystretch{0.8}
	\begin{center}
		\begin{tabular}{ccccccc}
			\hline\hline
			$\zeta$       & \multicolumn{2}{c}{Quarks}  & Leptons    & \multicolumn{3}{c}{Scalars}  \\\hline
			$-1/\sqrt{3}$ & $Q_{-1/3}(2)$ & $Q_{2/3}(1)$ & $L_{0}(3)$  & $H_S$ & $\eta^0$& $\eta^\pm$  \\\hline
			$+1/\sqrt{3}$ & $Q_{-1/3}(1)$ & $Q_{2/3}(2)$ & $L_{-1}(3)$ & $H_S$ & $\eta^0$& $\eta^\pm$  \\\hline
			$-\sqrt{3}$   & $Q_{-4/3}(2)$ & $Q_{5/3}(1)$ & - & $H_S$   & $\eta^\pm$& $\eta^{\pm\pm}$  \\\hline
			$+\sqrt{3}$   & $Q_{-4/3}(1)$ & $Q_{5/3}(2)$ & $L_{-2}(3)$ & $H_S$& $\eta^\pm$& $\eta^{\pm\pm}$  \\\hline\hline
		\end{tabular}
		\caption{Content of heavy extra degrees of freedom in the Class-I models with four $\zeta$ values. 
The subscript with quarks ($Q$) and leptons ($L$) represents the electric charge, 
and the number inside parentheses denotes the number of flavors. 
For scalar particles, $H_S$ represents a $Z_2^{\rm rem}$-even real scalar boson, while $\eta^0$, $\eta^\pm$ and $\eta^{\pm\pm}$
are $Z_2^{\rm rem}$-odd complex neutral, singly-charged and doubly-charged scalar bosons, respectively. 
                  }
		\label{table2}
	\end{center}
\end{table}

Before closing this section, let us comment on the other Class-I models with $\zeta = +1/\sqrt{3}$ and $\zeta = \pm \sqrt{3}$. 
In these models, heavy extra degrees of freedom, which are decoupled by taking the large $v_3$ limit, 
are different from those in the model with $\zeta = -1/\sqrt{3}$. 
In Table~\ref{table2}, we show the content of extra fermions and scalar bosons decoupled from the theory. See also Appendix~\ref{sec:app2} for details of the particle content. 
The important thing here is that when we take the large $v_3$ limit, 
the effective theory of these three models is the same as that of the model with $\zeta = -1/\sqrt{3}$. 
For the Class-II models, however, the effective theory is different from that discussed in this section, because all the three generations of quarks are embedded into 
the same $SU(3)_L$ representation, which do not lead to the flavor dependent structure as shown in Eq.~(\ref{eq:eff}).  
In addition, the lepton sector involves more complicated structures including a sextet representation in the flipped model and nice triplets in the $E_6$ model, so that 
the lepton sector in their effective theory can be richer than that discussed in this section.

\section{Flavor Constraints \label{sec:flavor}}

As we have seen in Sec.~\ref{sec:models}, 3-3-1 models introduce a 2HDM with a characteristic flavor structure in quark Yukawa interactions at the EW scale.
We thus take into account flavor constraints on the parameter space, particularly those from neutral meson mixings, $B\to X_s\gamma$ and EDMs. 
There are the other flavor constraints associated with leptons such as $M^\pm \to \ell^\pm\nu$, $M^0 \to \ell^+\ell^-$ and $\tau^\pm \to M^\pm \nu$
with $M^\pm$ and $M^0$ being charged and neutral mesons, respectively.
These processes typically become important in the case where leptonic Yukawa interactions are largely enhanced, see e.g., Refs.~\cite{Crivellin:2013wna,Arbey:2017gmh,Enomoto:2015wbn}.  
In our model, such an enhancement can be realized for large $\tan\beta$ cases which are highly constrained by the processes considered in this section. 
We thus do not discuss such processes in details. 

\subsection{Meson mixings \label{sec:meson}} 

We first consider the constraint from meson mixings which happen at tree level via neutral Higgs boson exchanges, because of the flavor violating interactions given in Eq.~(\ref{eq:gamma0}). 
According to Ref.~\cite{Gabbiani:1996hi}, new contributions to the $B_q^0$--$\bar{B}_q^0$ ($q=d,s$) mixing $\Delta m_{B_q}$ are given by 
\begin{align}
\Delta m_{B_q} &=2\text{Re}\left[C_{LR}^{B_q}\left(\frac{1}{24} + \frac{m_{B_q}^2}{4(m_b+m_q)^2}\right)-\frac{5}{24}\left(C_{LL}^{B_q} + C_{RR}^{B_q}\right)\frac{m_{B_q}^2}{(m_b+m_q)^2}\right]m_{B_q}f_{B_q}^2, 
\end{align}
where $m_{B_q}$ and $f_{B_q}$ are the mass of the $B_q^0$ meson and the decay constant, respectively. \footnote{The coefficients $C_{LL}$, $C_{RR}$ and $C_{LR}$ respectively correspond to those for the 
effective operators $Q_2$, $\tilde{Q}_2$ and $Q_4$ defined in Ref.~\cite{Gabbiani:1996hi}. }
The expressions for the $D^0$--$\bar{D}^0$ and $K^0$--$\bar{K}^0$ mixings are also obtained by the replacement of 
$(m_{B_q}^{},f_{B_q}^{},m_b,m_q,C_{ij}^{B_q}) \to (m_D^{},f_D^{},m_u,m_c,C_{ij}^K)$ and 
$(m_{B_q}^{},f_{B_q}^{},m_b,m_q,C_{ij}^{B_q}) \to (m_K^{},f_K^{},m_d,m_s,C_{ij}^K)$, respectively. 
The coefficients $C_{ij}^{M}$ $(M=B_q,D,K)$ are expressed as:
\begin{align}
\begin{split}
C_{LL}^{B_q}&=-\sum_\varphi\frac{\big[p_\varphi^d m_q(\Gamma^{d*}_{\varphi})_{3q}\big]^{2}}{2m^2_\varphi v^2} ,~	
C_{RR}^{B_q}=-\sum_\varphi\frac{\big[p_\varphi^d m_b(\Gamma^d_{\varphi })_{q3}\big]^{2}}{2m^2_\varphi v^2},~	
C_{LR}^{B_q}=-\sum_\varphi\frac{m_qm_b[(\Gamma^d_{\varphi} )_{q3}]^2}{m^2_\varphi v^2} , \\
C_{LL}^D&=-\sum_\varphi\frac{\big[p_\varphi^u m_u(\Gamma^{u*}_{\varphi})_{21}\big]^{2}}{2m^2_\varphi v^2} ,~	
C_{RR}^D=-\sum_\varphi\frac{\big[p_\varphi^u m_c(\Gamma^u_{\varphi })_{12}\big]^{2}}{2m^2_\varphi v^2},~	
C_{LR}^D=-\sum_\varphi\frac{m_um_c[(\Gamma^u_{\varphi} )_{12}]^2}{m^2_\varphi v^2} , \\
C_{LL}^K&=-\sum_\varphi\frac{\big[p_\varphi^d m_d(\Gamma^{d*}_{\varphi})_{21}\big]^{2}}{2m^2_\varphi v^2} ,~	
C_{RR}^K=-\sum_\varphi\frac{\big[p_\varphi^d m_s(\Gamma^d_{\varphi })_{12}\big]^{2}}{2m^2_\varphi v^2},~ 	
C_{LR}^K=-\sum_\varphi\frac{m_dm_s[(\Gamma^d_{\varphi} )_{12}]^2}{m^2_\varphi v^2}, 
\end{split}
\end{align}
where the summation takes over $\varphi=\{h,H,A\}$. 
For $M = B_q$, $(\Gamma^{d*}_{\varphi})_{3q}$ represents $(\Gamma^{d*}_{\varphi})_{31}$ and $(\Gamma^{d*}_{\varphi})_{32}$ for $q=d$ and $q=s$, respectively, 
and the same thing applies to $(\Gamma^{d}_{\varphi})_{q3}$.
We checked that these expressions are consistent with those given in Refs.~\cite{Crivellin:2013wna,Jana:2021tlx}.

At the alignment limit, i.e., $s_{\beta-\alpha} = 1$, the contribution from the $h$ mediation vanishes, because the Yukawa coupling of $h$ takes a diagonal form. 
When we take $m_H = m_A$ in addition to $s_{\beta-\alpha} = 1$, $C_{LL}$ and $C_{RR}$ vanish, because the cancellation happens between the contributions from the $H$ and $A$ mediations. 
In these limits, the expression for $\Delta m_M$ ($M=B,D,K$) takes a simple form as
\begin{align}
\begin{split}
\Delta m_K &=m_K\frac{m_dm_sf_B^2}{m^2_A v^2}\left[\frac{1}{6} + \frac{m_K^2}{(m_d+m_s)^2}\right]\text{Re}\left\{[(\Gamma^d_A)_{12}]^2\right\}, \\
\Delta m_D &=m_D\frac{m_um_cf_D^2}{m^2_A v^2}\left[\frac{1}{6} + \frac{m_D^2}{(m_u+m_c)^2}\right]\text{Re}\left\{[(\Gamma^u_A)_{12}]^2\right\}, \\ 
\Delta m_{B_q} &=m_{B_q}\frac{m_qm_bf_B^2}{m^2_A v^2}\left[\frac{1}{6} + \frac{m_{B_q}^2}{(m_q + m_b)^2}\right]\text{Re}\left\{[(\Gamma^d_A)_{q3}]^2\right\}. 
\end{split}\label{bbtheo}
\end{align}
The current measured values of $\Delta m_M$ are~\cite{Zyla:2020zbs}
\begin{align}
\begin{split}
\Delta m_K^{\rm exp}    & = (3.484 \pm 0.006)\times 10^{-12}~~\text{MeV},\\
\Delta m_D^{\rm exp}    & = (6.251 \pm 2.70)\times 10^{-12}~~\text{MeV}, \\
\Delta m_{B_d}^{\rm exp} & = (3.334 \pm 0.013)\times 10^{-10}~~\text{MeV}, \\
\Delta m_{B_s}^{\rm exp} & = (1.1677 \pm 0.0013)\times 10^{-8}~~\text{MeV}. 
\end{split}\label{bbexp}
\end{align}
By comparing Eqs.~(\ref{bbtheo}) and (\ref{bbexp}), we can constrain $m_A(=m_H^{})$, $t_\beta$ and the mixing/phase parameters ($\theta$, $\psi$, $\delta_{4,5,6}$). 
In the numerical evaluation, we simply require that the absolute values of the new contribution given in Eq.~(\ref{bbtheo}) do not exceed the central value of the measurement. 

\subsection{$B \to X_s\gamma$ \label{sec:bsg}}

The mass of the charged Higgs boson and its Yukawa coupling are constrained by the $B \to X_s \gamma$ process. 
The current measured value of the branching ratio is given by~\cite{HFLAV:2019otj} 
\begin{align}
  {\cal B}(B \to X_s\gamma) = (3.32\pm  0.15)\times 10^{-4}. 
\end{align}
The bound on $m_{H^\pm}$ has been studied in the 2HDMs with a softly-broken $Z_2$ symmetry at the next-to-leading order (NLO)~\cite{Ciuchini:1997xe,Kagan:1998ym,Borzumati:1998tg,Borzumati:1998nx} and at the next-to-NLO (NNLO)~\cite{Misiak:2006zs,Misiak:2017bgg,Misiak:2020vlo} in QCD, where $m_{H^\pm} \gtrsim$ 800 GeV has been taken in the Type-II 2HDM with $t_\beta \gtrsim 2$~\cite{Misiak:2020vlo}. 

In our effective 2HDM, the situation is similar to the Type-II 2HDM if the mixing parameters $\theta$, $\psi$ and the phases $\delta_{4,5,6}$
are taken to be small. However, when these parameters get larger values, the constraint should be drastically different from the Type-II case. 
In fact, we will see in Sec.~\ref{sec:num} that the bound from the $B \to X_s \gamma$ decay can be relaxed when the mixing parameters are switched on. 

For the numerical evaluation of the branching ratio of $B \to X_s \gamma$, we employ the expression given in Ref.~\cite{Borzumati:1998tg}, in which
QCD and QED corrections are implemented at NLO. 
In Ref.~\cite{Borzumati:1998tg}, the branching ratio is given in terms of the $X$ and $Y$ parameters which are the coefficients of the Yukawa coupling for the charged Higgs boson\footnote{In the 
Type-I (Type-II) 2HDM, $X=-\cot\beta~(\tan\beta)$ and $Y = \cot\beta~(\cot\beta)$~\cite{Borzumati:1998tg}. }, and 
they appear as combinations of $XY^*$ and $|Y|^2$. 
In our model, $XY^*$ and $|Y|^2$ are expressed as\footnote{We only consider the dominant top loop contribution to the decay rate. }
\begin{align}
XY^* &= \frac{(M_L^*)_{33}(M_R)_{23}}{m_tm_b(V_{\rm CKM})_{33}(V_{\rm CKM}^*)_{32}},\quad 
|Y|^2 = \frac{(M_R^*)_{33}(M_R)_{23}}{m_t^2(V_{\rm CKM})_{33}(V_{\rm CKM}^*)_{32}}. 
\end{align}

\subsection{Electric Dipole Moments \label{sec:edm}}

\begin{figure}
\centering
\includegraphics[width=150mm]{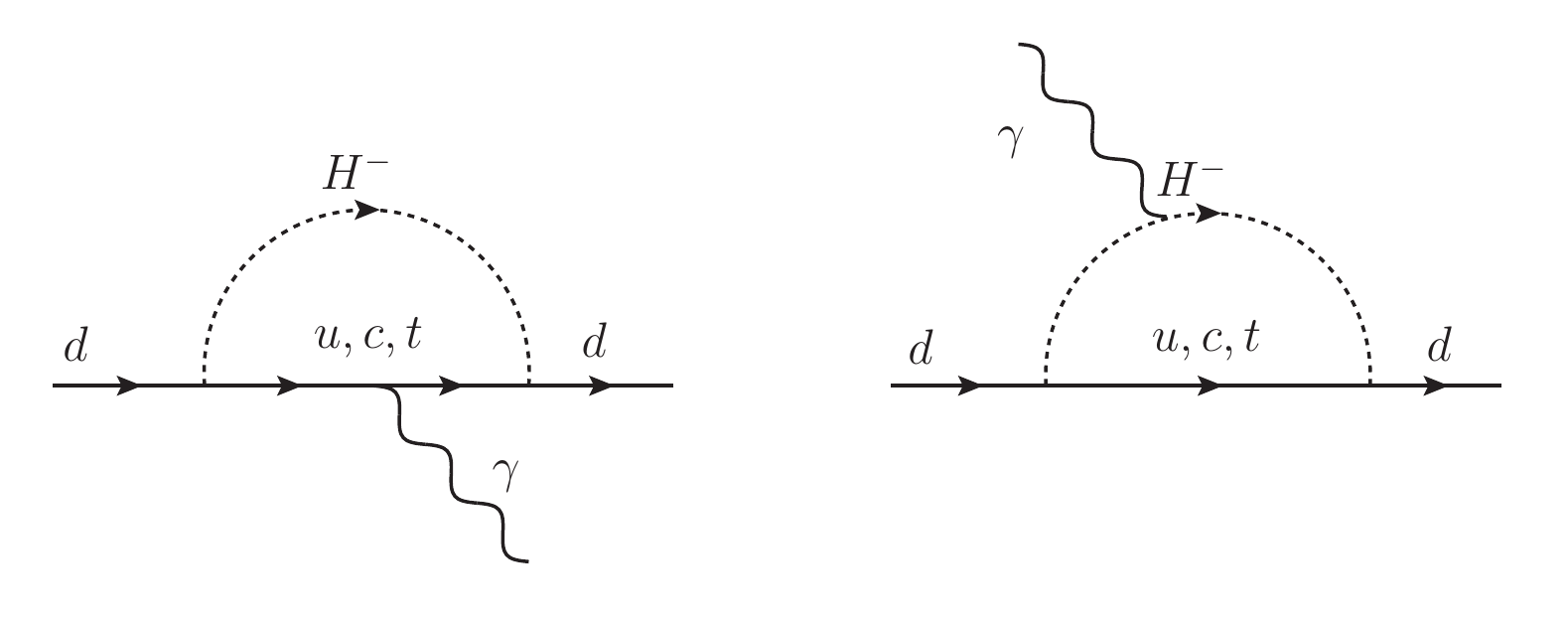}
\caption{One-loop contributions to $d_d$.  }
\label{fig:edm}
\end{figure}

Generally, new CPV phases in the Yukawa interaction can induce new contributions to EDMs. 
Effects of CPV can be described by the following effective Lagrangian up to dimension six as  
\begin{align}
	\mathcal{L}_{\text{eff}} &= -\frac{id_f}{2}\overline{\psi}_f\sigma^{\mu\nu}\gamma_5\psi_fF_{\mu\nu}-
	\frac{id^C_f}{2}\overline{\psi}_f\sigma^{\mu\nu}\gamma_5T^A\psi_fG^A_{\mu\nu} \notag\\
	& +\frac{1}{3}C_Wf_{ABC}G_{\mu \nu}^A  \tilde{G}^{B\,\nu\rho} G_{\rho}^{C\,\mu}
	+ C_{ff'} (\bar{f}f)(\bar{f}' i\gamma_5 f').  \label{eq:edm}
\end{align}
In the first line, $d_f$ $(d_f^C)$ denotes the coefficients of the EDM and the Chromo EDM (CEDM),  
where $\sigma^{\mu\nu} = \frac{i}{2}[\gamma^\mu,\gamma^\nu]$, $F_{\mu\nu}$ and $G_{\mu\nu}^A$ are respectively the field strength tensor for the photon and gluon with $A (=1,\dots,8)$ being the index for the $SU(3)_C$ adjoint representation. 
In the second line, $C_W$ ($C_{ff'}$) represents the coefficients of the Weinberg operator~\cite{Weinberg:1989tt} (four-fermi interaction), where 
$f_{ABC}$ is the structure constant of the $SU(3)$ group and $ \tilde{G}^A_{\mu\nu} = \epsilon_{\mu\nu\rho\sigma}G^{A\,\rho\sigma}$. 

We first consider one-loop contributions to the (C)EDM. 
The contributions from the neutral Higgs boson $\varphi=\{h,H,A\}$ loops are calculated to be zero shown as 
\begin{align}
d_q \propto \text{Im}[\Gamma_\varphi^q\Gamma_\varphi^q]_{11} = 
\text{Im}[(\Gamma_\varphi^q)_{1i}(\Gamma_\varphi^q)_{i1}]
=\text{Im}[(\Gamma_\varphi^q)_{1i}(\Gamma_\varphi^{q*})_{1i}]
=\text{Im}[|(\Gamma_\varphi^q)_{1i}|^2] = 0. 
\end{align}
Here, we used the hermiticity of the matrices $\Gamma_\varphi^q$.
The contribution from the charged Higgs boson loop, shown in Fig.~\ref{fig:edm}, is expressed as 
\begin{align}
&\hspace{-5mm} d_d^{\text{1-loop}}= \frac{em_d}{8\pi^2v^2}\text{Im}\left[
Q_u\left(V_{\rm CKM}^\dagger \Gamma_A^u\, G_1^u \,  \bar{\Gamma}_A^u V_{\rm CKM}\right)_{11}+\left(V_{\rm CKM}^\dagger \Gamma_A^u \, G_2^u \, \bar{\Gamma}_A^u V_{\rm CKM}  \right)_{11}\right], \label{eq:1loop1}\\
&\hspace{-5mm} d_u^{\text{1-loop}} = \frac{em_u}{8\pi^2v^2}\text{Im}\left[Q_d\left(\bar{\Gamma}_A^u V_{\rm CKM} G_1^d V_{\rm CKM}^\dagger \Gamma_A^u \right)_{11}-\left(\bar{\Gamma}_A^u V_{\rm CKM}\, G_2^d\, V_{\rm CKM}^\dagger \Gamma_A^u \right)_{11}\right], \label{eq:1loop2}
\end{align}
where $Q_u~(Q_d) = 2/3~(-1/3)$, and we use $\Gamma_A^d = V_{\rm CKM}^\dagger \bar{\Gamma}_A^u V_{\rm CKM}$ 
with $\bar{\Gamma}_A^u \equiv \Gamma_A^u|_{\text{diag}(-t_\beta,-t_\beta,t_\beta^{-1}) \to \text{diag}(t_\beta^{-1},t_\beta^{-1},-t_\beta)}$.
The matrices $G_{1,2}^{q}$ are defined as 
\begin{align}
G_i^u &= \text{diag}\left[G_i\left(\frac{m_{H^\pm}^2}{m_u^2}\right),G_i\left(\frac{m_{H^\pm}^2}{m_c^2}\right),G_i\left(\frac{m_{H^\pm}^2}{m_t^2} \right)\right],\\
G_i^d &= \text{diag}\left[G_i\left(\frac{m_{H^\pm}^2}{m_d^2}\right),G_i\left(\frac{m_{H^\pm}^2}{m_s^2}\right),G_i\left(\frac{m_{H^\pm}^2}{m_b^2} \right)\right]~~(i=1,2), 
\end{align}
with 
\begin{align}
G_1(x) = \frac{1-4x+3x^2-2x^2\ln x}{2(1-x)^3}, \quad 
G_2(x) = \frac{1-x^2 + 2x\ln x}{2(1-x)^3}. 
\end{align}
Similarly, the contribution to the CEDM is obtained by replacing the external photon with the external gluon in the left diagram of Fig.~\ref{fig:edm}. We obtain 
\begin{align}
d_d^{C\text{1-loop}} &= \frac{g_sm_d}{8\pi^2v^2}\text{Im}\left[V_{\rm CKM}^\dagger \Gamma_A^u\, G_1^u\, \bar{\Gamma}_A^u V_{\rm CKM}  \right]_{11},~ 
d_u^{C\text{1-loop}} = \frac{g_sm_u}{8\pi^2v^2}\text{Im}\left[ \bar{\Gamma}_A^u V_{\rm CKM}\, G_1^d\,   V_{\rm CKM}^\dagger \Gamma_A^u \right]_{11}, \label{eq:1loop2c}
\end{align}
where $g_s$ is the $SU(3)_C$ gauge coupling constant.

The magnitude of $d_d^{(C)\text{1-loop}}$ is mainly determined by the top-loop diagram, because the other contributions are suppressed by $m_q^2/m_{H^\pm}^2$ ($q \neq t$). 
Thus, the one-loop contribution can approximately be expressed as 
\begin{align}
d_d^\text{1-loop} &\simeq 
 \frac{em_d}{24\pi^2v^2}
\left[2G_1\left(\frac{m^2_{H^\pm}}{m^2_t}\right) + 3G_2\left(\frac{m^2_{H^\pm}}{m^2_t}\right)\right]
\text{Im}\left[\left(V_{\rm CKM}^\dagger \Gamma_A^u\right)_{13} \left(\bar{\Gamma}_A^u V_{\rm CKM}  \right)_{31}\right], \label{eq:app1}  \\
d_d^{C\text{1-loop}} &\simeq 
 \frac{g_sm_d}{8\pi^2v^2}
G_1\left(\frac{m^2_{H^\pm}}{m^2_t}\right)\text{Im}\left[\left(V_{\rm CKM}^\dagger \Gamma_A^u\right)_{13} 
\left(\bar{\Gamma}_A^u V_{\rm CKM}\right)_{31}\right].\label{eq:app2}
\end{align}
We note that $d_u^{(C)\text{1-loop}}$ is negligibly smaller than $d_d^{(C)\text{1-loop}}$ because of the absence of such heavy quark loops. 

It is important to emphasize here that one-loop contributions to $d_q$ are negligibly smaller than two-loop Barr-Zee type contributions
in models with flavor conservation at tree level such as the four types of the 2HDMs and Yukawa aligned 2HDMs. 
This is because one-loop contributions from charged Higgs boson loops to $d_q$ are significantly suppressed  
by the facotr of $m_d\times m_U^2/m_{H^\pm}^2|V_{\rm CKM}^{Ud}|^2$ ($U=u,c,t$) for $q=d$ and $m_u\times m_D^2/m_{H^\pm}^2|V_{\rm CKM}^{uD}|^2$ ($D=d,s,b$) for $q=u$, while 
those from neutral Higgs boson loops vanish. 
On the other hand, the two-loop Barr-Zee contribution with a neutral Higgs and a photon exchange 
is only linearly suppressed by a small fermion mass and not suppressed by a tiny CKM matrix element, 
so that it usually dominates with respect to the one-loop contribution. 
The above statement does not hold in models without flavor conservation at tree level as in our present model.

\begin{figure}
	\centering
	\includegraphics[scale=0.8]{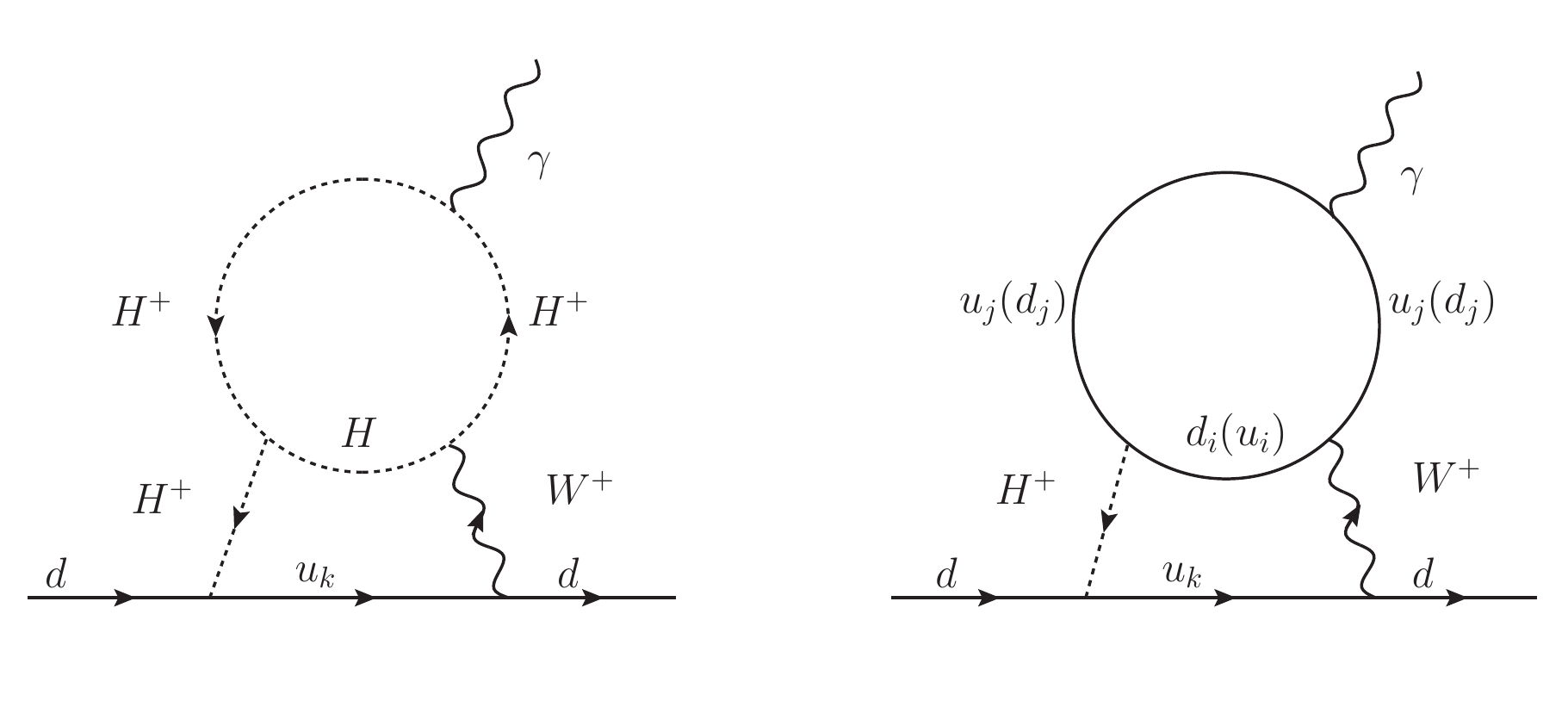}
	\caption{Contributions to $d_d$ from two-loop Barr-Zee type diagrams. }
        \label{fig:bz}
\end{figure} 

In order to show the dominance of the one-loop contribution, we consider contributions from two-loop Barr-Zee type diagrams~\cite{Barr:1990tg} to $d_q$. 
Similar to the one-loop case, diagrams with neutral Higgs boson exchanges do not contribute
to the EDM, because of the hermiticity of the matrices $\Gamma_\varphi^q$. 
Thus, the diagram with the charged Higgs boson exchange, shown in Fig.~\ref{fig:bz}, contributes to the EDM. 
In the alignment limit $s_{\beta-\alpha}\to 1$, contributions from $W$-$W$-$h$ and $W$-$W$-$H$ loops vanish, and 
the scalar-loop $d_d^{\text{BZ}}(S)$ and the fermion-loop $d_d^{\text{BZ}}(F)$ give non-zero contributions. 
These are calculated as
\begin{align}
d_d^{\text{BZ}}(S) &= -\left(\frac{1}{16\pi^2}\right)^2\frac{eg^2}{2v}\lambda_{H^+ H^- H}\sum_k\text{Im}(M_L^{1k} V_{\rm CKM}^{k1})\notag\\
&\hspace{-10mm}\times \int_0^1dx x
\left[C_0^{\rm BZ}(\bar{m}_{H^\pm}^2 + \tilde{m}_{H}^2, m_W^2,m_{H^\pm}^2) +  m_{u_k}^2D_0^{\rm BZ}(\bar{m}_{H^\pm}^2 + \tilde{m}_{H}^2, m_W^2, m_{u_k}^2,m_{H^\pm}^2)\right], \\
d_d^{\text{BZ}}(F) &= -\left(\frac{1}{16\pi^2} \right)^2 \frac{eg^2}{v^2}N_c  \int_0^1 dx\, \left(Q_d + \frac{1-x}{x}Q_u \right) \notag\\
&  \times \sum_{i,j,k}\text{Im} \left[M_L^{1k}V_{\rm CKM}^{k1} \left(x\,M_L^* V_{\rm CKM}^*M_d^{\rm diag} + (x-2)\,M_R^* M_u^{\rm diag} V_{\rm CKM}^* \right)_{ij} \right]  \notag\\
& \times   \left[C_0^{\rm BZ}(\bar{m}_{d_i}^2 + \tilde{m}_{u_j}^2, m_W^2,m_{H^\pm}^2) +  m_{u_k}^2D_0^{\rm BZ}(\bar{m}_{d_i}^2 + \tilde{m}_{u_j}^2, m_W^2,m_{u_k}^2,m_{H^\pm}^2 )\right], 
\end{align}
where $\bar{m}_X^2 = m_X^2/(1-x)$, $\tilde{m}_X^2 = m_X^2/x$, 
and $M_{L,R}$ are given in Eq.~(\ref{eq:mlmr}). 
The scalar trilinear coupling $\lambda_{H^+H^-H}$ is defined by the coefficient of the vertex $H^+H^-H$ in the Lagrangian, which is extracted to be 
\begin{align}
\lambda_{H^+H^-H} = \frac{2}{v}(m_H^2 - m_A^2)\cot 2\beta , 
\end{align}
at the alignment limit. 
The loop functions $C_0^{\rm BZ}$ and $D_0^{\rm BZ}$  are given by 
\begin{align}
C_0^{\rm BZ}(a,b,c)& = -\frac{a\ln a}{(a-b)(a-c)} + \text{(cyclic)} , \\
D_0^{\rm BZ}(a,b,c,d) &= \frac{a\ln a}{(a-b)(a-c)(a-d)} + \text{(cyclic)}, 
\end{align}
where (cyclic) denotes the other terms obtained by replacing the first term in the cyclic way, e.g., $(a,b,c)\to (b,c,a)$. 
Similarly, we obtain 
\begin{align}
d_u^{\text{BZ}}(S) &=  \left(\frac{1}{16\pi^2}\right)^2\frac{eg^2}{2v}\lambda_{H^+ H^- H}\sum_k\text{Im}(V_{\rm CKM}^{1k}M_R^{k1})\notag\\
&   \hspace{-10mm}\times \int_0^1dx x
\left[C_0^{\rm BZ}(\bar{m}_{H^\pm}^2 + \tilde{m}_{H}^2, m_W^2,m_{H^\pm}^2) +  m_{d_k}^2D_0^{\rm BZ}(\bar{m}_{H^\pm}^2 + \tilde{m}_{H}^2, m_W^2, m_{d_k}^2,m_{H^\pm}^2)\right], \\
d_u^{\text{BZ}}(F) &= -\left(\frac{1}{16\pi^2} \right)^2 \frac{eg^2}{v^2} N_c \sum_{i,j,k} \int_0^1 dx
\left(Q_u + \frac{x}{1-x}Q_d \right) \notag\\
&\times \sum_{i,j,k}\text{Im}\left[V_{\rm CKM}^{1k}M_R^{k1}\left((1+x)M_L^* V_{\rm CKM}^*M_d^{\rm diag} - (1-x)M_R^* M_u^{\rm diag} V_{\rm CKM}^* \right)_{ij} \right]\notag\\
&\times \left[C_0^{\rm BZ}(\bar{m}_{d_i}^2 + \tilde{m}_{u_j}^2, m_W^2,m_{H^\pm}^2) +  m_{d_k}^2D_0^{\rm BZ}(\bar{m}_{d_i}^2 + \tilde{m}_{u_j}^2, m_W^2,m_{d_k}^2,m_{H^\pm}^2 )\right]. 
\end{align}
We note that the diagrams which are obtained by replacing the external photon by the gluon in Fig.~\ref{fig:bz} vanish, because the color index is not closed in the loop, so that the Barr-Zee type diagram does not contribute to $d_q^C$. 

The nEDM $d_n$ can be calculated by a linear combination of the EDM and CEDM by using the QCD sum rule, which are expressed as~\cite{Abe:2013qla} 
\begin{align}
d_n = 0.79d_d - 0.20d_u + \frac{e}{g_s}(0.59d_d^C + 0.30 d_u^C). 
\end{align}
In our calculation, we identify $d_q = d_q^{\text{1-loop}} + d_q^{\text{BZ}}$ and $d_q^C = d_q^{C\text{1-loop}}$. 
The current upper limit on the nEDM $d_n$ has been given by
\begin{align}
|d_n|<1.8\times 10^{-26}~e\, \text{cm}~~ (90\%~~\text{CL}), \label{eq:nedm-exp}
\end{align}
from the nEDM experiment~\cite{nEDM:2020crw}. 
In future, the bound on $|d_n|$ will be improved to be $1\times 10^{-27} e \, \text{cm}$ by the n2EDM experiment with its
designed performance and to be of order $10^{-28}~e\,\text{cm}$ by its possible modifications~\cite{n2EDM:2021yah}. 

We here comment on the other contributions to $d_n$ from the Weinberg operator $C_W$ and the four-fermi interaction $C_{ff}$. 
It has been known that the contribution from the charged Higgs boson exchange in two-loop diagrams to $C_W$ vanishes~\cite{Jung:2013hka,Logan:2020mdz}. 
In addition, the contribution from the neutral Higgs boson exchanges to $C_W$ in two-loop diagrams and that to $C_{ff}$ in tree level diagrams
vanish due to the hermicity of the $\Gamma_\varphi^q$ matrices. 

Let us also comment on the constraint from the electron EDM $d_e$ whose current upper limit is given to be $|d_e|< 1.1\times 10^{-29}e\,\text{cm}$ at 90\% CL at the ACME experiment~\cite{ACME:2018yjb}.
\footnote{The ACME experiment has been performed with thorium monoxide (ThO), which has a sensitivity to a combination of $d_e$ and $C_S$ with the latter being 
related to the coefficient of the electron-nucleon interaction $(\bar{N}N)(\bar{e}i\gamma_5 e)$. 
The value of $C_S$ can be estimated from $C_{ff'}$ which are zero at tree level in our model as mentioned above, while 
the bound $|d_e|< 1.1\times 10^{-29}e\,\text{cm}$ has been obtained by taking $C_S$ to be zero~\cite{ACME:2018yjb}. }
\footnote{In addition to the EDM measured through paramagnetic systems such as ThO, that from diamagnetic atoms, e.g., Hg~\cite{Griffith:2009zz} and Xe~\cite{Inoue:2013fxa} 
also provide a sensitivity to the coefficients given in Eq.~(\ref{eq:edm}). It is, however, quite difficult to obtain robust constraints on the coefficients from them due to 
huge theoretical uncertainties~\cite{Jung:2013hka}. }
The dominant contribution to $d_e$ comes from Barr-Zee diagrams which are obtained by replacing the external down quark $d$ (internal up-type quarks $u_k$) 
with the electron (neutrinos) in Fig.~\ref{fig:bz}. 
Since there is no CPV phase in the lepton Yukawa couplings for $H^\pm$ in the limit of massless neutrinos, only the diagram with quark loops (right panel of Fig.~\ref{fig:bz}) contributes to $d_e$. 
Thus, $d_e$ can be expressed as 
\begin{align}
d_e &\simeq  -\left(\frac{1}{16\pi^2} \right)^2 \frac{eg^2m_e}{v^2}t_\beta N_c  \int_0^1 dx\, \left(Q_d + \frac{1-x}{x}Q_u \right) \notag\\
&  \times \sum_{i,j,k}\text{Im} \left[\left(x\,M_L^* V_{\rm CKM}^*M_d^{\rm diag} + (x-2)\,M_R^* M_u^{\rm diag} V_{\rm CKM}^* \right)_{ij} \right]  C_0^{\rm BZ}(\bar{m}_{d_i}^2 + \tilde{m}_{u_j}^2, m_W^2,m_{H^\pm}^2). 
\end{align}
We confirm that the value of $|d_e|$ can be maximally about $7 \times 10^{-31}\,e$\,cm for $m_{H^\pm} = 600$ GeV and $t_\beta = 3$ by scanning the parameters $(\theta,\psi,\delta_{4,5,6})$, 
so that we can safely avoid the current upper limit. 
For $t_\beta > 10$, $|d_e|$ can exceed the upper limit, but such a large $t_\beta$ value is also excluded by the constraints from $d_n$ as well as the flavor experiments as we will see in the next section. 

\section{Numerical evaluations \label{sec:num}}

\begin{table}[h]
	\begin{center}
		\begin{tabular}{llllllll}
			\hline
			\hline
$\alpha_{\rm em}^{-1}$~\cite{Zyla:2020zbs}&$m_u$~\cite{Xing:2007fb}& $m_c$~\cite{Xing:2007fb} & $m_d$~\cite{Xing:2007fb} & $m_s$~\cite{Xing:2007fb} & $m_b$~\cite{Xing:2007fb} & $m_B^{}$~\cite{Zyla:2020zbs}\\
137.036 & 1.9 MeV & 0.901 GeV  &4.22 MeV & 80 MeV & $4.2$ GeV & 5.279 GeV \\\hline
$m_{B_s}^{}$~\cite{Zyla:2020zbs} & $m_D^{}$~\cite{Zyla:2020zbs} & $m_K^{}$~\cite{Zyla:2020zbs}  & $f_B^{}$~\cite{FlavourLatticeAveragingGroup:2019iem} &$f_{B_s}^{}$~\cite{FlavourLatticeAveragingGroup:2019iem}
& $f_D^{}$ ~\cite{FlavourLatticeAveragingGroup:2019iem}& $f_K^{}$~\cite{FlavourLatticeAveragingGroup:2019iem}  \\   
5.366 GeV&1.865 GeV&0.498 GeV&0.190 GeV&0.230 GeV&0.212 GeV&0.156 GeV\\
			\hline
			\hline
		\end{tabular}
\caption{SM input parameters for the calculation of the meson mixings and ${\cal B}(B\to X_s\gamma$), where we take $m_{B_d} = m_B$ and  $f_{B_d} = f_B$. }
\label{table3}
	\end{center}
%
	\begin{center}
		\begin{tabular}{cccccccc}
			\hline
			\hline
$\alpha_{\rm em}^{-1}$~\cite{Zyla:2020zbs} & $\alpha_{s}$~\cite{FlavourLatticeAveragingGroup:2019iem} & $m_u$~\cite{Xing:2007fb} & $m_c$~\cite{Xing:2007fb}  & $m_d$~\cite{Xing:2007fb} & $m_s$~\cite{Xing:2007fb} & $m_b$~\cite{Xing:2007fb}  \\\hline
127.952&  0.1182 &$1.27$ MeV&$0.619$ GeV&$2.93$ MeV& $55$ MeV&$2.89$ GeV\\
			\hline
			\hline
		\end{tabular}
		\caption{SM input parameters at the $m_Z^{}$ scale for the calculation of the EDM constraints. }
                \label{table4}
	\end{center}
%
	\begin{center}
		\begin{tabular}{cccccccccc}
			\hline
			\hline
$m_t$~\cite{Xing:2007fb} & $m_Z^{}$~\cite{Zyla:2020zbs} & $m_W^{}$~\cite{Zyla:2020zbs} & $A$~\cite{Charles:2015gya}  & $\lambda$~\cite{Charles:2015gya} & $\overline{\rho}$~\cite{Charles:2015gya} & $\overline{\eta}$~\cite{Charles:2015gya}\\
			\hline
$171.7$ GeV& 91.1876 GeV& 80.379 GeV& 0.810  & 0.22548 & 0.145 & 0.343 \\
			\hline
			\hline
		\end{tabular}
		\caption{Common SM input parameters.}
                \label{table5}
	\end{center}
\end{table}

In this section, we numerically evaluate the constraint on the parameter space from the meson mixings, the $B \to X_s\gamma$ decay and the nEDM $d_n$ discussed in Sec.~\ref{sec:flavor}. 
We then study decays of the additional Higgs bosons in a few benchmark points which are allowed by all the constraints considered in this section. 

We use the SM input parameters summarized in Tables~\ref{table3}, \ref{table4} and \ref{table5}. 
In Table~\ref{table5}, $A$, $\lambda$, $\bar{\rho}$ and $\bar{\eta}$ denote the Wolfenstein parameters for the CKM matrix defined as~\cite{ParticleDataGroup:2020ssz} :
\begin{align}
	\label{39}
	V_{\text{CKM}}=\left(
	\begin{matrix}
		1-\lambda^2/2& \lambda  & A \lambda ^3 (\bar{\rho} -i \bar{\eta}) \\
		-\lambda  & 1-\lambda ^2/2 & A \lambda ^2 \\
		A \lambda ^3 (1 -\bar{\rho} -i \bar{\eta}) & -A \lambda ^2 & 1 \\
	\end{matrix}
	\right)+\mathcal{O}(\lambda^4). 
\end{align}

In the following, we particularly focus on the case with the alignment limit $s_{\beta-\alpha} \to 1$ for simplicity, in which the $h$ state has the same couplings as those of the SM Higgs boson at tree level, although 
a small deviation from the alignment limit is allowed by the current LHC data~\cite{ATLAS:2021vrm,CMS:2020gsy}, e.g., $|c_{\beta-\alpha}| \lesssim 0.1$ at $\tan\beta \simeq 1$ in the Type-II 2HDM. 
Such a deviation can mainly affect on the constraint from the meson mixings as discussed in Sec.~\ref{sec:meson}, while its influence on the EDMs can be negligible, because the hermiticity of the $\Gamma_\varphi^q$ matrices
also works for the non-alignment case. 
For the impact of the deviation from the alignment limit on the meson mixings, see Ref.~\cite{Okada:2016whh}. 

We also would like to mention constraints on the masses of the additional Higgs bosons from experiments other than flavor ones. 
First, they can be constrained from the EW oblique parameters such as the $S$, $T$ and $U$ parameters~\cite{Peskin:1990zt}. 
In particular, the $T$ parameter severely constrains the mass difference between charged and neutral Higgs bosons, 
because it breaks the custodial $SU(2)$ symmetry~\cite{Pomarol:1993mu}. 
It has been known that new contributions to the $T$ parameters vanish in the case with $m_A = m_{H^\pm}$, and also those to $S$ disappear for $m_H = m_A = m_{H^\pm}$, see e.g., Refs.~\cite{Kanemura:2011sj,Blasi:2017zel}.
We thus take the degenerate mass case ($m_H = m_A = m_{H^\pm}$) in what follows, in which the constraints from the meson mixings also become quite milder than the case with $m_H \neq m_A$. 
The mass of the additional Higgs bosons can also be constrained from direct searches for additional Higgs bosons at LHC. 
As we will see in Sec.~\ref{sec:decay}, 
their decay products typically include top quarks e.g., $H/A \to t\bar{u},~t\bar{c}$, $t\bar{t}$ and $H^\pm \to tb$. 
The current LHC data do not exclude scenarios with additional Higgs bosons which dominantly decay into the above modes~\cite{Hou:2020chc}.

\subsection{Case without new phases}

We first discuss the case without the new CPV phases, i.e., $\delta_{4,5,6}=0$, in which 
the CPV effect purely comes from the CKM phase.  
As we will see at the end of this subsection, the contribution from the two-loop Barr-Zee type diagrams to $d_n$ is negligibly smaller than the one-loop contribution. 
Thus, we focus on the one-loop contribution. 

Using the Wolfenstein parameterization given in Eq.~(\ref{39}), the approximate formulae for $d_d^{(C)\text{1--loop}}$ given in Eqs.~(\ref{eq:app1}) and (\ref{eq:app2}) 
are expressed up to ${\cal O}(\lambda^5)$ as 
\begin{align}
\frac{d_d^{\text{1--loop}}}{e} &\simeq \frac{A \bar{\eta}\lambda^3}{24\pi^2}\frac{m_d(1-t_\beta^2)}{v^2s_{\beta}^2}
c_\theta c_\psi\left(s_{\theta} -\lambda c_\theta s_\psi - \frac{\lambda^2}{2} s_{\theta} \right)\left[2G_1\left(\frac{m^2_{H^\pm}}{m^2_t}\right) +3G_2\left(\frac{m^2_{H^\pm}}{m^2_t}\right) \right], \label{47}\\
\frac{d^{C\text{1--loop}}_{d}}{g_s} &\simeq \frac{A \bar{\eta}\lambda^3}{8\pi^2}\frac{m_d(1-t_\beta^2)}{v^2s_{\beta}^2}c_\theta c_\psi\left(s_{\theta} -\lambda c_\theta s_\psi - \frac{\lambda^2}{2} s_{\theta} \right)G_1\left(\frac{m^2_{H^\pm}}{m^2_t}\right). \label{48}
\end{align}
As we can see that the dominant contribution is proportional to $\lambda^3$, because one has to 
pick up the CPV parameter $\bar{\eta}$ in the element $V_{\rm CKM}^{13}$ or $V_{\rm CKM}^{31}$ at least once. 
Before numerical evaluations, let us give a few remarks on the above approximate formulae as follows: 
\begin{enumerate}
\item For $\psi = \pi/2 + n\pi$ $(n \in \mathbb{Z})$, $d_d^{(C)\text{1--loop}}$ vanishes up to ${\cal O}(\lambda^5)$ 
as it is proportional to $c_\psi$. In this case, 
the matrices $\Gamma_A^u$ and $\bar{\Gamma}_A^u$ are a block diagonal form with $2\times 2$ and $1\times 1$ blocks, 
so that the product of the combinations $(V_{\rm CKM}^\dagger \Gamma_A^u)_{13} [= (V_{\rm CKM}^*)_{31} (\Gamma_A^u)_{33}]$ and
$ (\bar{\Gamma}_A^u V_{\rm CKM})_{31} [= (\bar{\Gamma}_A^u)_{33}(V_{\rm CKM})_{31}]$ gives a real value at order $\lambda^6$. 
\item For $\theta = \pi/2 + n\pi$ $(n \in \mathbb{Z})$, $d_d^{(C)\text{1--loop}}$ vanishes at all orders of $\lambda$. 
In this case, the matrices $\Gamma_A^u$ and $\bar{\Gamma}_A^u$ are diagonal forms, and the combinations 
$(V_{\rm CKM}^\dagger \Gamma_A^u\, G_i^u \,  \bar{\Gamma}_A^u V_{\rm CKM})$ and $(\bar{\Gamma}_A^u V_{\rm CKM} G_1^d V_{\rm CKM}^\dagger \Gamma_A^u)$
appearing in Eqs.~(\ref{eq:1loop1}) and (\ref{eq:1loop2}) are hermitian matrices.
\item For $t_\beta = 1$, $d_d^{(C)\text{1--loop}}$ vanishes at all orders of $\lambda$. In this case, 
$\bar{\Gamma}_A^u = -\Gamma_A^u$ holds, so that $(V_{\rm CKM}^\dagger \Gamma_A^u\, G_i^u \,  \bar{\Gamma}_A^u V_{\rm CKM})$ and $(\bar{\Gamma}_A^u V_{\rm CKM} G_1^d V_{\rm CKM}^\dagger \Gamma_A^u)$ are hermitian matrices.
\item $|d_d^{(C)\text{1--loop}}|$ is centrosymmetric about $(\theta,\psi)=(n\pi/2,m\pi)$ ($n,m \in \mathbb{Z}$)
and symmetric about the lines with $\psi=\pi/2+n\pi$ $(n \in \mathbb{Z})$.
\end{enumerate}

\begin{figure}
  \centering
  \includegraphics[width=75mm]{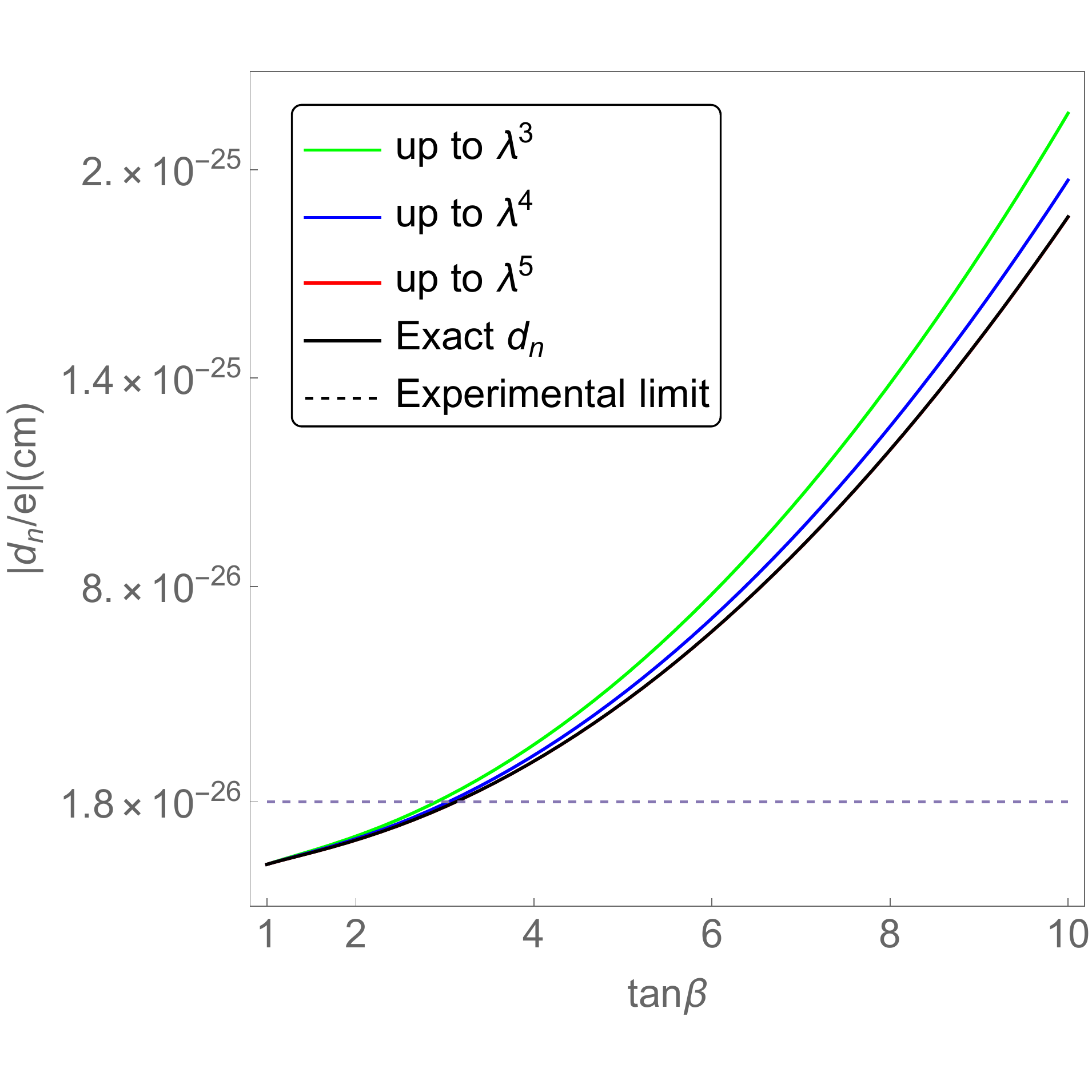}
  \includegraphics[width=75mm]{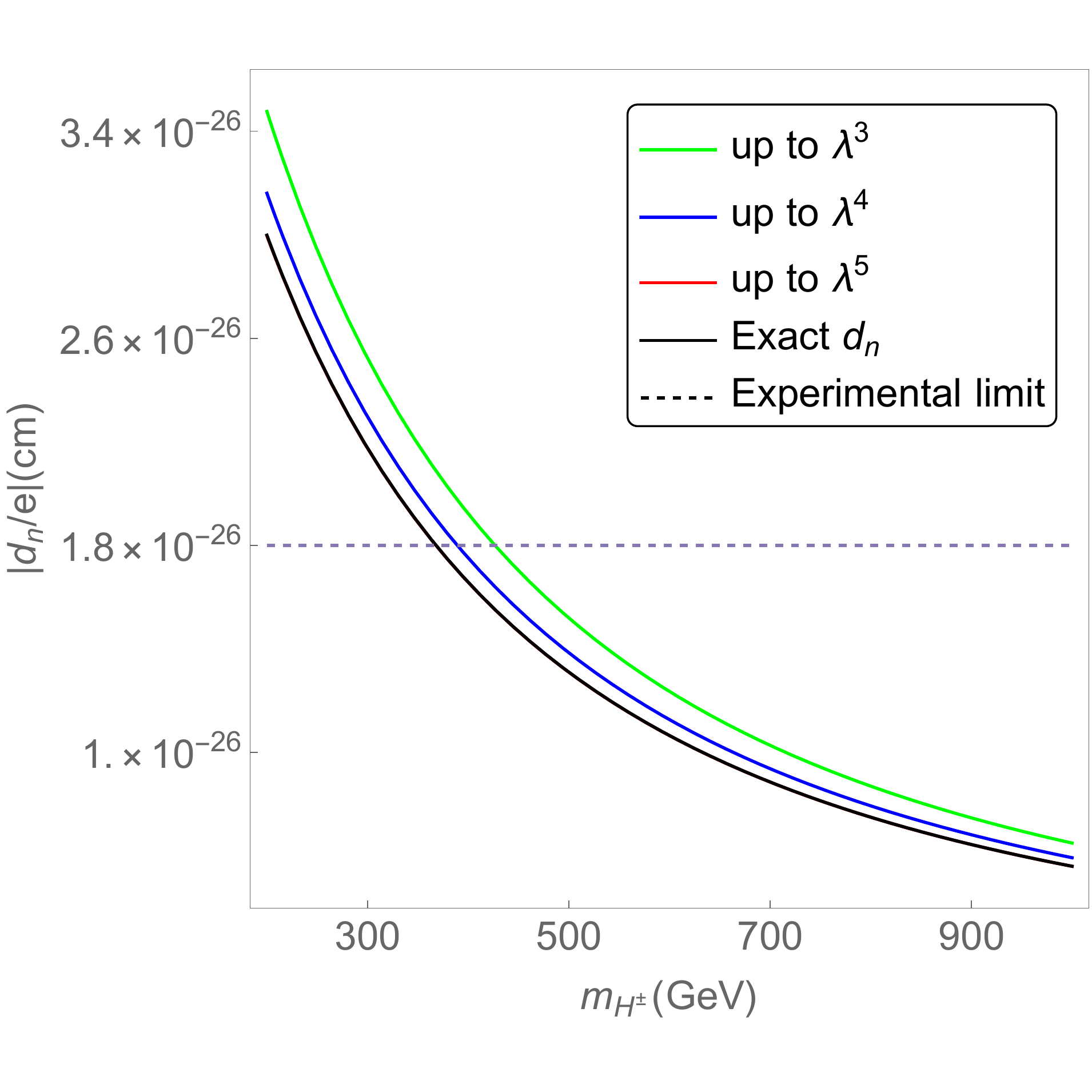}
  \caption{Comparison of the value of $|d_n/e|$ for $\theta=\psi=\pi/3$ and $\delta_{4,5,6}=0$ by using the approximate formulae given in Eqs.~(\ref{47}) and (\ref{48}) up to $\mathcal{O}(\lambda^3)$, $\mathcal{O}(\lambda^4)$ and $\mathcal{O}(\lambda^5)$ and 
by using the exact formulae given in Eqs.~(\ref{eq:1loop1}), (\ref{eq:1loop2}) and (\ref{eq:1loop2c}). 
We fix  $m_{H^\pm}=400$ GeV (left) and $\tan\beta =3$ (right).}
  \label{fig2}
\end{figure}

In Fig.~\ref{fig2}, we show the comparison of the numerical values of $d_n$ at one-loop level 
by using the exact formula given in Eqs.~(\ref{eq:1loop1}), (\ref{eq:1loop2}) and (\ref{eq:1loop2c}) 
and the approximate one given in Eqs.~(\ref{47}) and (\ref{48}) as a function of $t_\beta$ (left panel) and $m_{H^\pm}$ (right panel). 
For the approximate formula, we use the expression up to $\mathcal{O}(\lambda^3)$, $\mathcal{O}(\lambda^4)$ and $\mathcal{O}(\lambda^5)$. 
We here fix the mixing angles $\theta = \psi = \pi/3$ as an example, 
but we confirm that the validity of the approximation is not spoiled depending on values of the mixing angles except for the case with $\psi = \pi/2 + n\pi$ ($n \in \mathbb{Z}$)~\footnote{In this case, 
the approximation is no longer good, because the contribution from ${\cal O}(\lambda^6)$ turns out to be dominant. 
It, however, gives negligibly smaller values of $|d_n|$ as compared with the current upper limit, so that we do not need to care about such breakdown of the approximation. }. 
It is clear that the approximation with the $\mathcal{O}(\lambda^5)$ term is quite good agreement with the exact one.  
In addition, we see that $|d_n|$ increases by $t_\beta^2$ for $t_\beta > 1$, 
and decreases by $m_t^2/m_{H^\pm}^2$ for $m_{H^\pm} \gg m_t$ as they can be read from Eqs.~(\ref{47}) and (\ref{48}).  

\begin{figure}
  \centering
  \includegraphics[width=80mm]{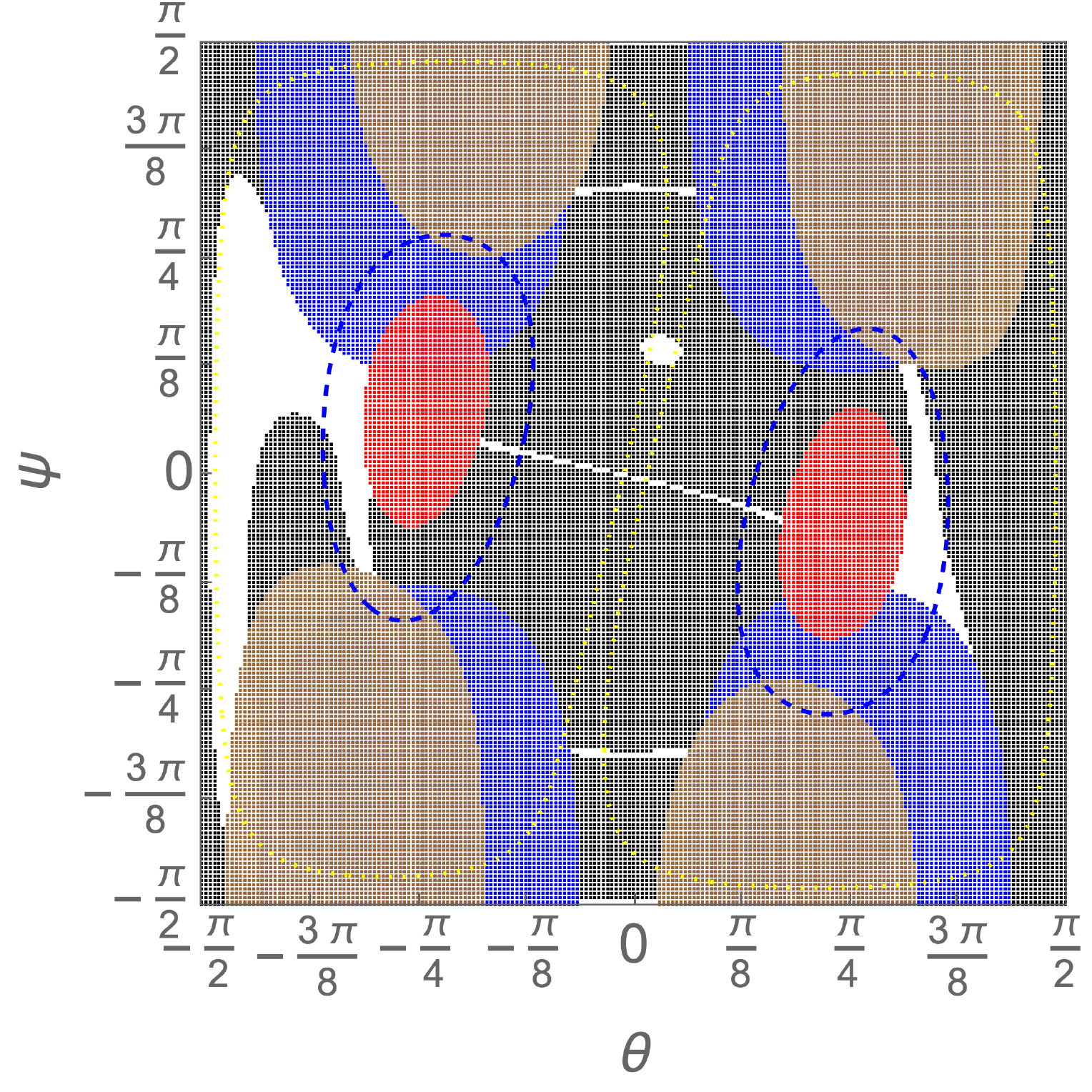}
  \includegraphics[width=80mm]{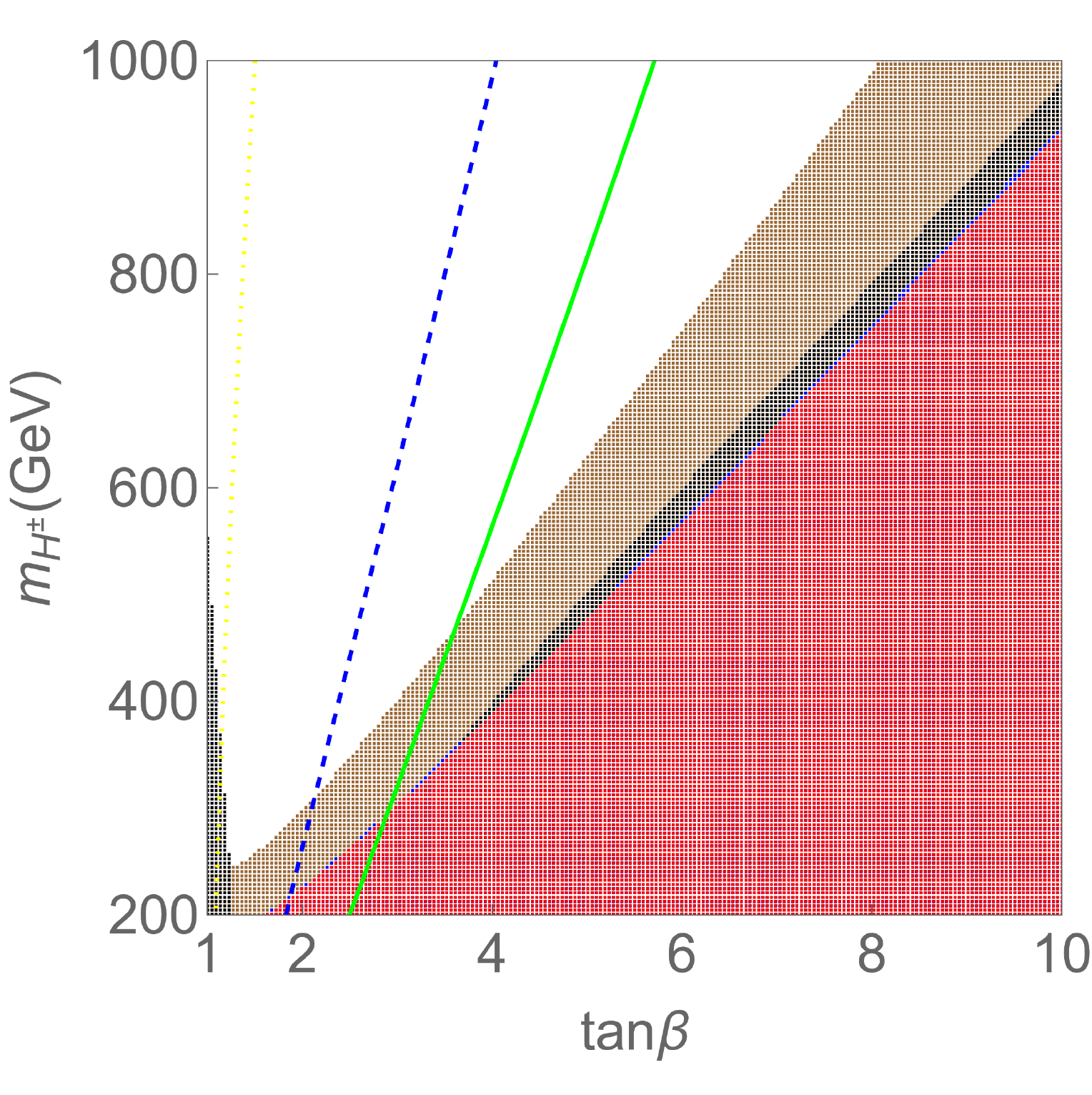}
  \caption{Contour plots of $|d_n|$ on the $\theta$--$\psi$ plane with $t_\beta=2$ and $m_{H^\pm}^{} = 600$ GeV (left panel) and $t_\beta$--$m_{H^\pm}^{}$ plane with $\theta=-7\pi/16$ and $\psi=\pi/8$ (right panel). 
We take the new phases to be zero ($\delta_{4,5,6} = 0$) and degenerate masses as $m_A^{} = m_H^{} = m_{H^\pm}^{}$. 
The shaded regions are excluded by the constraint from the $B^0$-$\bar{B}^0$ (red), $D^0$-$\bar{D}^0$ (blue), $K^0$-$\bar{K}^0$ (brown) and $B\to X_s \gamma$ (black) data, respectively. 
The green-solid, blue-dashed and yellow-dotted contours denote $|d_n| = 1.8\times 10^{-26}\, e \,\text{cm}$, $9.0\times 10^{-27}$ and $1.0\times 10^{-27}\, e \,\text{cm}$, respectively.  }
\label{fig:one-loop-num}
\end{figure}

Next in Fig.~\ref{fig:one-loop-num}, we show various parameter dependences, e.g., $\theta$, $\psi$, $t_\beta$ and $m_{H^\pm}$, of $|d_n|$ under the constraints from the meson mixings and $B\to X_s \gamma$ data. 
We numerically find that $|d_n|$ takes a maximum value at $(\theta,\psi) \simeq (-0.79,0.18)$, and these values almost do not depend 
on the choice of $t_\beta$ and $m_{H^\pm}$, as it is expected from Eqs.~(\ref{47}) and (\ref{48}). 
From the left panel of Fig.~\ref{fig:one-loop-num}, we see that the wide region of the parameter space is excluded by the flavor constraints, while 
the allowed region typically predicts values of $|d_n|$ with one order of magnitude smaller than the current upper limit, which can be explored at the n2EDM experiment. 
%
The right panel shows the constrained region on the $t_\beta$--$m_{H^\pm}^{}$ plane with the fixed values of the mixing angles $\theta=-7\pi/16$ and $\psi=\pi/8$, 
which are allowed by all the three constraints for $t_\beta = 2$ and $m_{H^\pm} = 600$ GeV, see the left panel. 
We can extract the upper limit on $t_\beta$ to be about 3 (5) for $m_{H^\pm}=400$ (800) GeV from the meson mixings ($|d_n|$).
Remarkably, the light charged Higgs boson with a mass of about 250 GeV is allowed depending on $\theta$, $\psi$ and $t_\beta$ by the $B\to X_s \gamma$ data. 
Such a light charged Higgs boson scenario has been excluded in the Type-II 2HDM. 

\begin{figure}
  \centering
  \includegraphics[width=80mm]{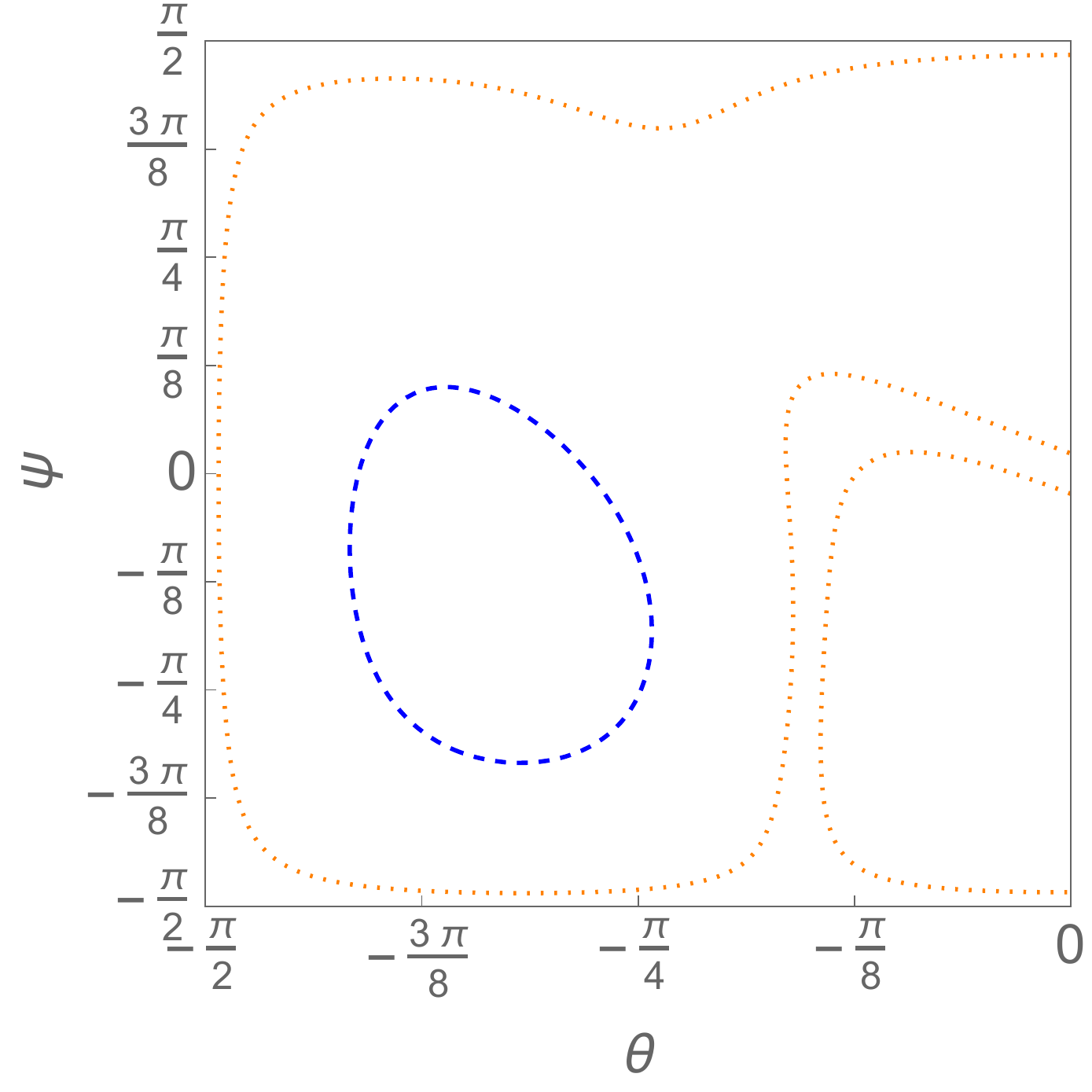}
  \includegraphics[width=80mm]{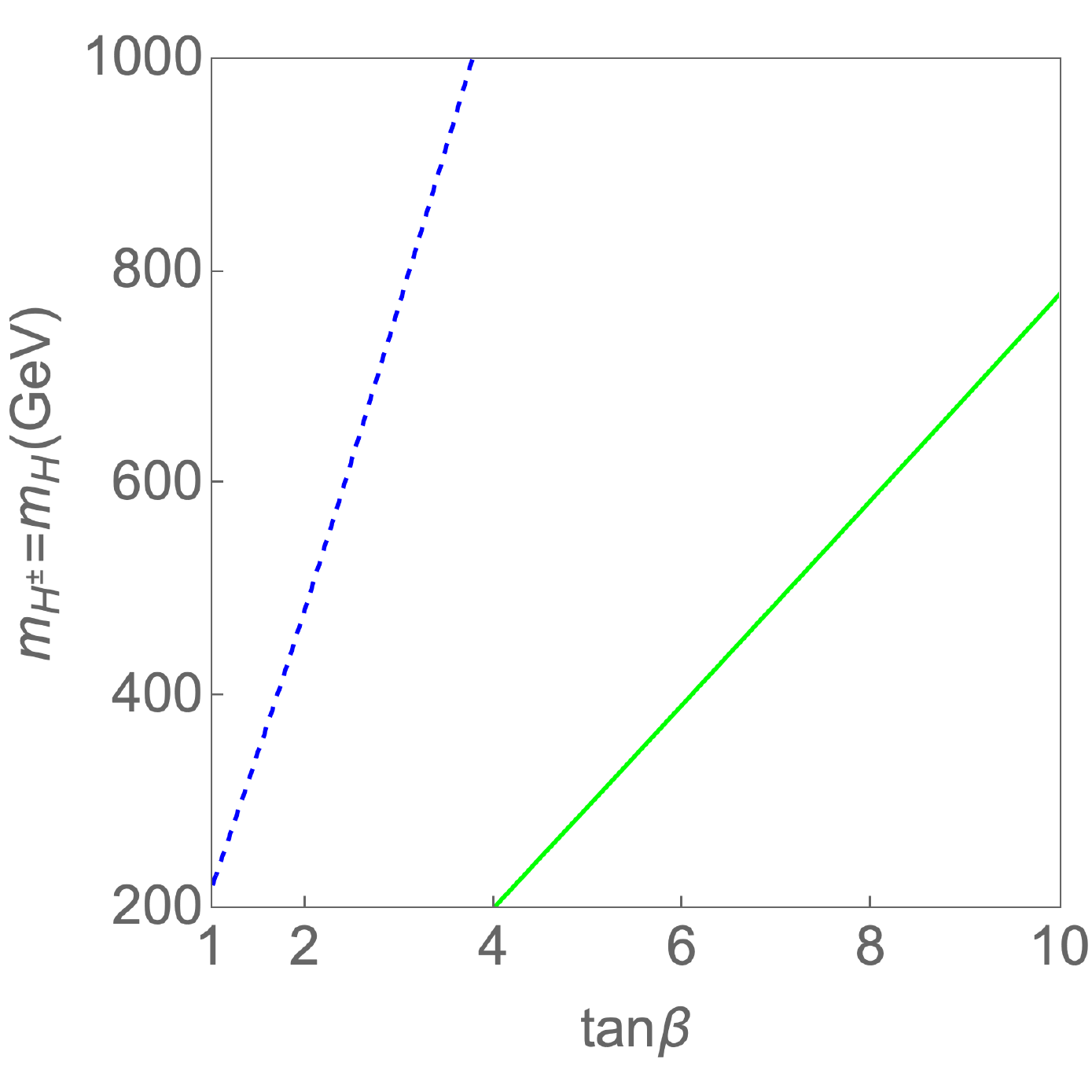}
 \caption{Contour plots of the contribution from two-loop Barr-Zee diagrams to $|d_n|$ on the $\theta$--$\psi$ plane 
with $t_\beta=2$ and $m_{H^\pm}^{} = 600$ GeV (left panel) and $t_\beta$--$m_{H^\pm}^{}$ plane with $\theta=-7\pi/16$ and $\psi=\pi/8$ (right panel). 
We take the new phases to be zero ($\delta_{4,5,6} = 0$) and degenerate masses as $m_A^{} = m_H^{} = m_{H^\pm}^{}$. 
The solid, dashed and dotted contours denote $|d_n| = 1.0\times 10^{-28}\, e \,\text{cm}$, $1.0\times 10^{-29}$ and $1.0\times 10^{-30}\, e \,\text{cm}$, respectively. 
}
\label{fig:barr-zee-num}
\end{figure}

As aforementioned, the contributions from the Barr-Zee diagram are negligibly smaller than the one-loop contribution which is shown in Fig.~\ref{fig:one-loop-num}. 
In order to numerically show this, we exhibit the value of $|d_n|$ from the Barr-Zee diagram only in Fig.~\ref{fig:barr-zee-num} with the same configuration used in Fig.~\ref{fig:one-loop-num}. 
From Fig.~\ref{fig:barr-zee-num}, we see that the contribution to $|d_n|$ is typically of order $10^{-29}$ or smaller, which is roughly two or more than two orders of magnitude smaller than the one-loop contribution. 
This behavior does not significantly change for the case with non-zero phases. 
We thus can safely neglect the Barr-Zee diagrams. 

\subsection{Non-zero phases}

\begin{figure}
	\centering
  \includegraphics[width=50mm]{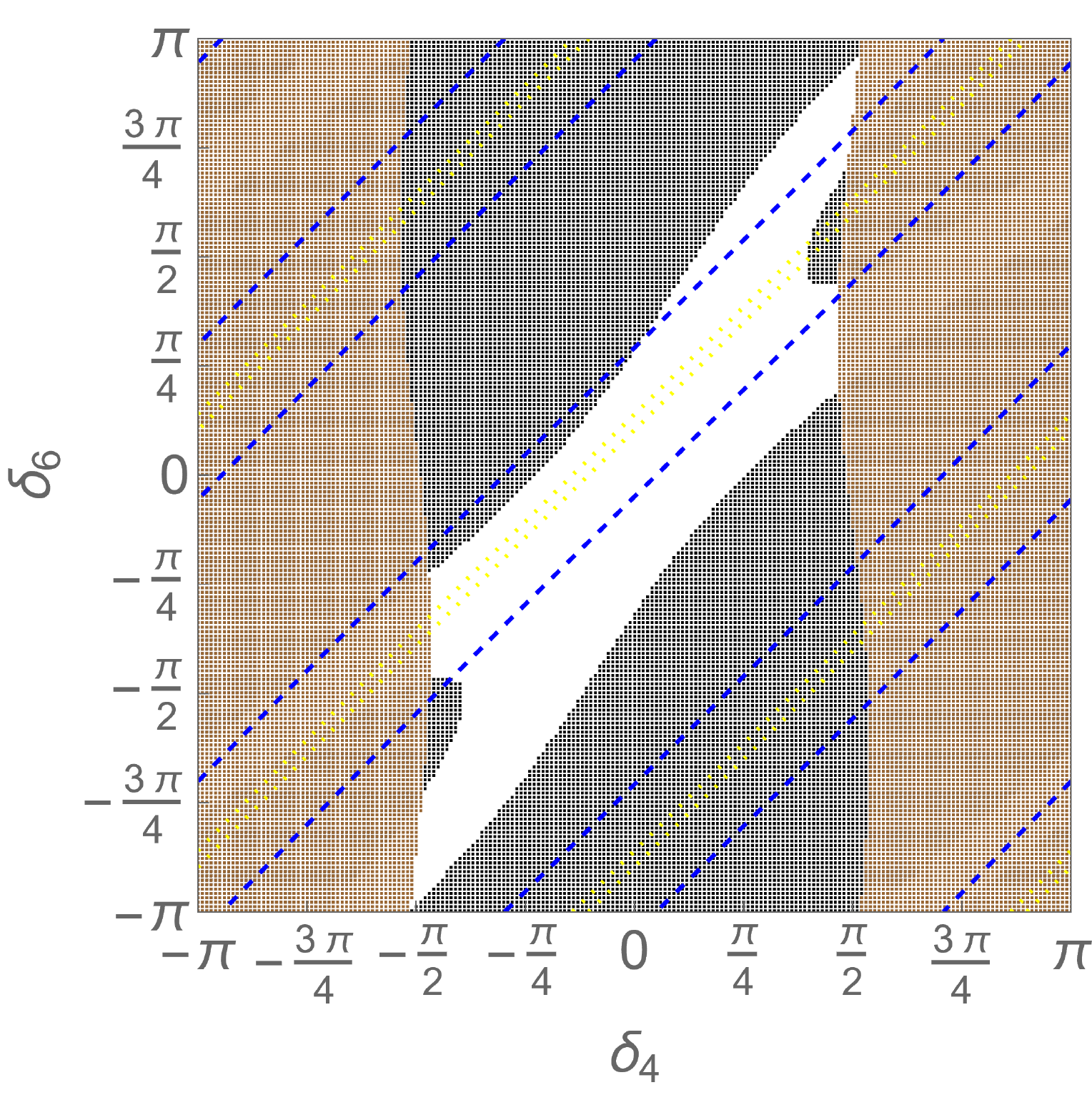}
  \includegraphics[width=50mm]{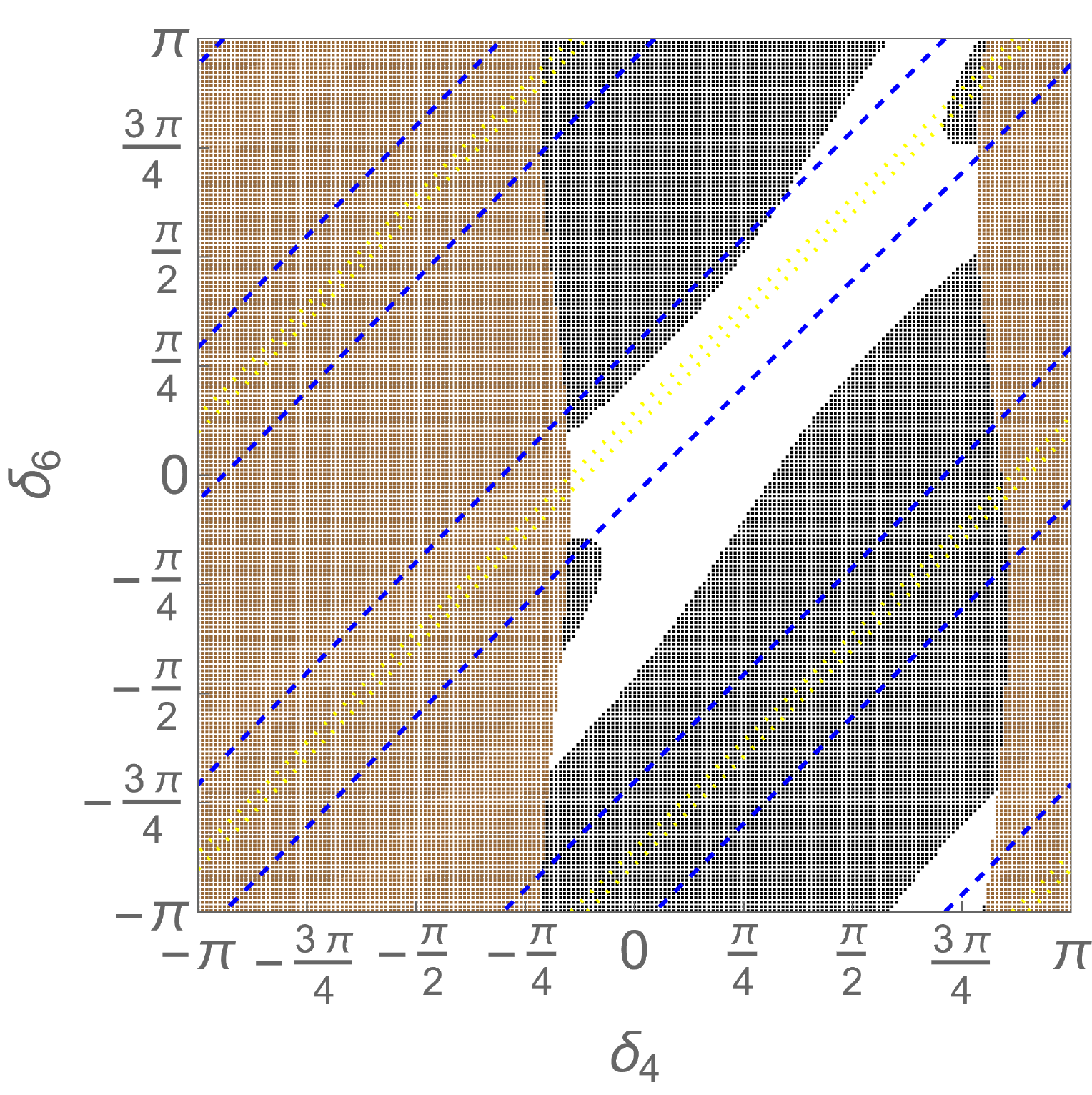}
  \includegraphics[width=50mm]{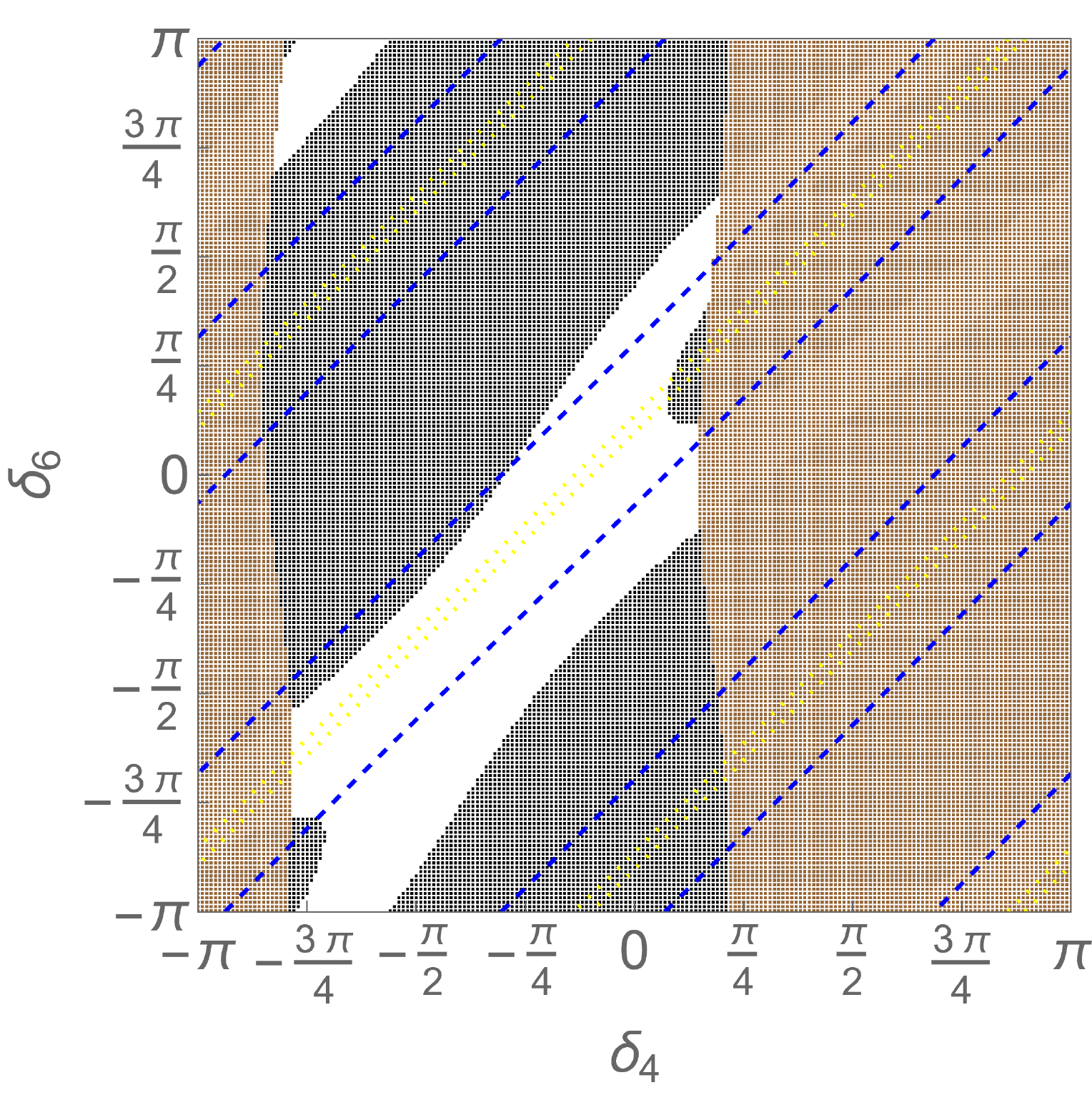}\\
  \includegraphics[width=50mm]{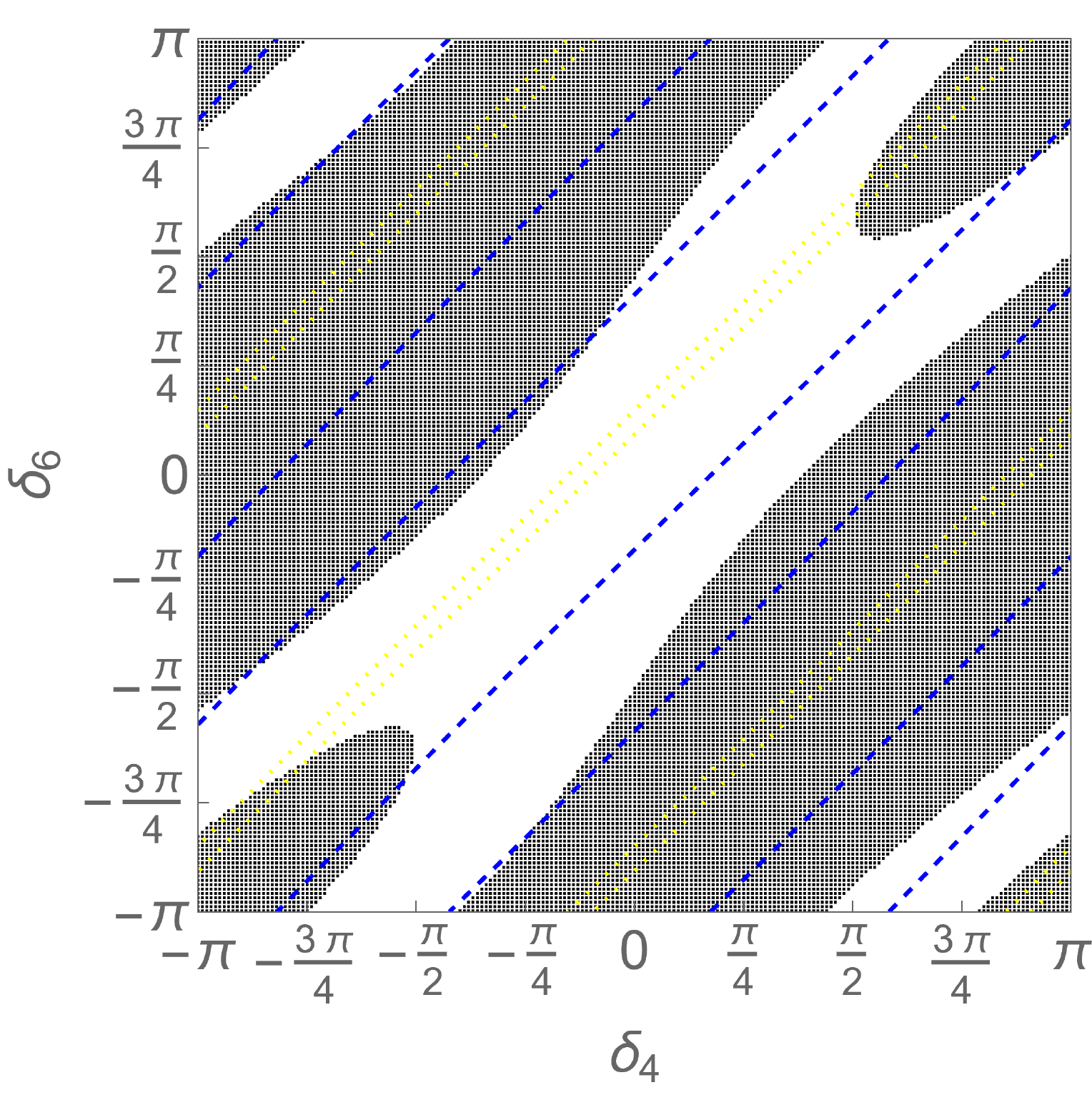}
  \includegraphics[width=50mm]{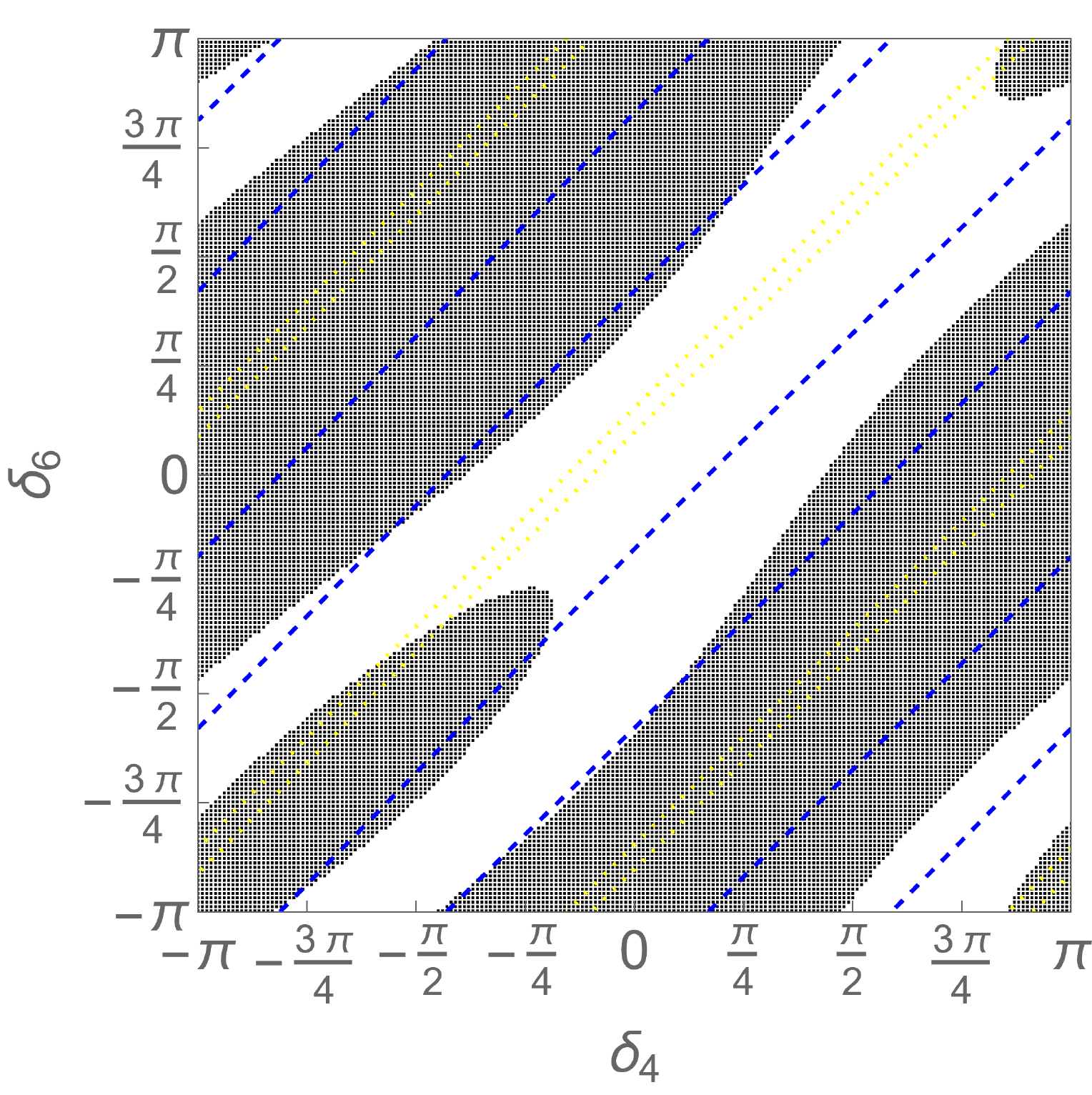}
  \includegraphics[width=50mm]{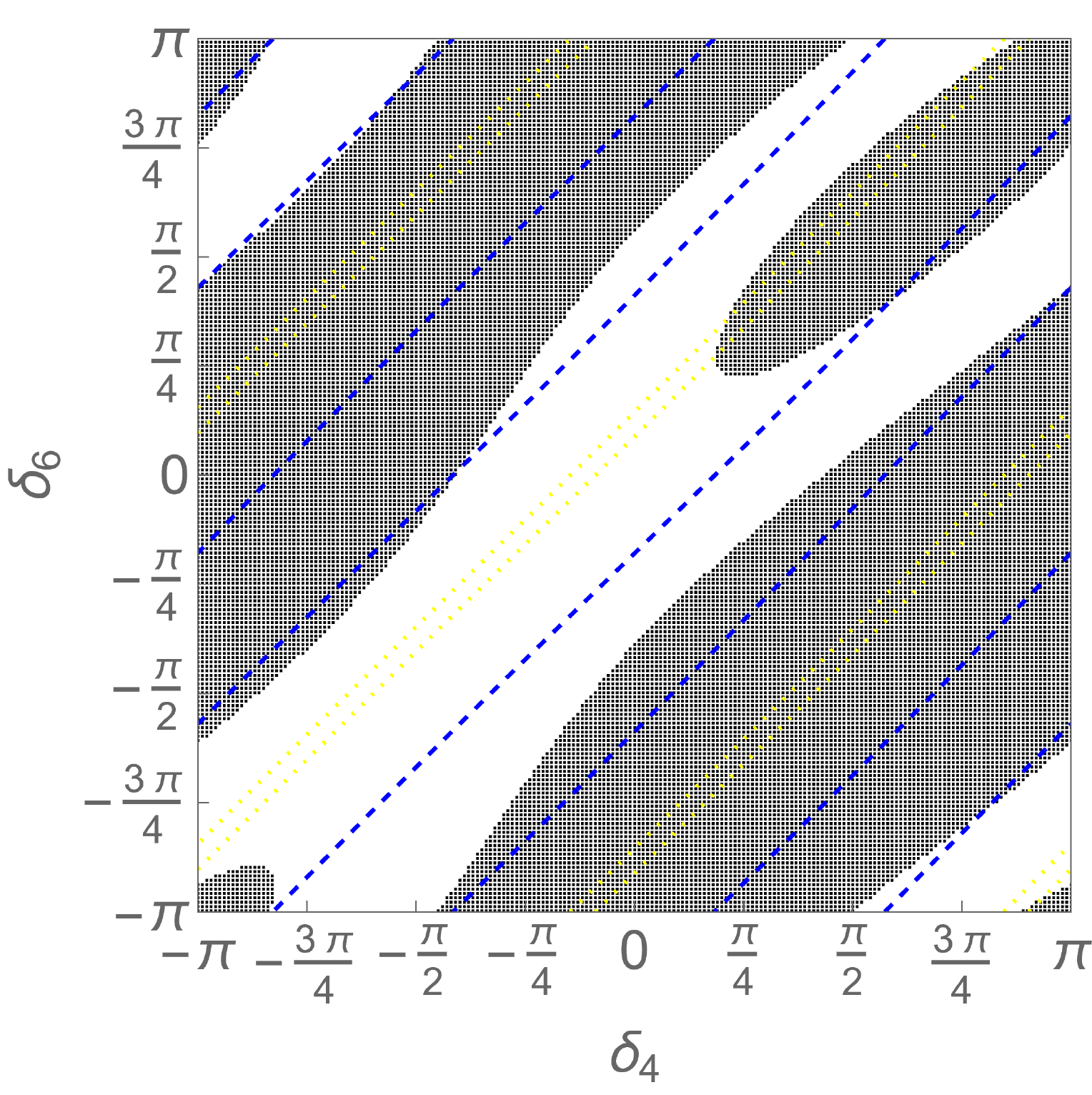}\\
\caption{Contour plots of $|d_n|$ with non-zero CPV phases $\delta_{4,5,6}$ on the $\delta_4$--$\delta_6$ plane with $t_\beta=2$, $m_H^{} = m_A^{} = m_{H^\pm}^{}$, $\theta=-7\pi/16$ and $\psi=\pi/8$. 
We take $m_{H^\pm} = 400$ GeV for the upper panels and 600 GeV for the lower panels. 
The remaining phase $\delta_5$ is taken to be  $\delta_5=0$ (left panel), $\delta_5=1$ (center panel) and $\delta_5=-1$ (right panel). 
The regions shaded by black color are excluded by the constraint from $B\to X_s \gamma$ data. 
The dashed and dotted contours denote $|d_n| = 9.0\times 10^{-27}$ and $1.0\times 10^{-27}\, e \,\text{cm}$, respectively. 
}
\label{fig:cpv}
\end{figure}

Next, we discuss the case with non-zero phases $\delta_{4,5,6}$. 
In Fig.~\ref{fig:cpv}, we show the region constrained from the meson mixings, $B\to X_s \gamma$ data and the nEDM
on the $\delta_4$--$\delta_6$ plane for $m_{H^\pm} = 400$ GeV (upper panels) and 600 GeV (lower panels). 
The other parameters are fixed as explicitly written in the caption. 
From the left panels ($\delta_5 = 0$), we see that at the point of the no phase limit $(\delta_4,\delta_6) = (0,0)$, 
the value of $|d_n|$ is given of order $10^{-27}$, while the region with $\delta_{4,6}\neq 0$ the value of $|d_n|$
can quickly glow up to $10^{-26}$. 
We also show the similar plots for $\delta_5 = 1$ (center) and $\delta_5 = -1$ (right), where the pattern of the prediction for $|d_n|$ is 
similar to the case with $\delta_5 = 0$, but the region allowed by the $B\to X_s \gamma$ data is shifted to upper and lower regions, respectively. 

To conclude, we clarify that the new CPV phases $\delta_{4,5,6}$ with $\mathcal{O}(1)$ are allowed by the current $B$ physics observables and the EDM data even for
the case with a relatively light charged Higgs boson with the mass of a few hundred GeV. 
It should be emphasized that our scenario becomes quite similar to the Type-II 2HDM in the no-mixing limit, i.e., $\theta = \psi = 0$, so that 
the constraint from the $B\to X_s \gamma$ decay is severe, i.e., $m_{H^\pm} \gtrsim 600$ GeV. 
This, however, can drastically be changed for the case with non-zero mixing and/or non-zero additional CPV phases, where 
the constraint from the $B\to X_s \gamma$ decay is significantly relaxed which makes a light charged Higgs boson scenario possible. 
Such a scenario is also possible in the Type-I or Type-X 2HDM, but our scenario can be distinguished from them by measuring 
flavor violating decays of the additional Higgs bosons, e.g., $H/A \to tc$ and/or 
flavor dependent deviations in the $h$ couplings from the SM predictions which do not appear in the $Z_2$ symmetric 2HDMs, see Ref.~\cite{Okada:2016whh}. 
In addition, our scenario can be probed by future EDM experiments, e.g., n2EDM, and also can directly be tested at collider experiments by looking at decays
of additional Higgs bosons, see e.g.,~\cite{Kanemura:2021atq} for the probe of CPV phases at electron-positron colliders.

\subsection{Decays of the additional Higgs bosons \label{sec:decay}}

\begin{table}
	\begin{center}
		\begin{tabular}{cccccccc}
			\hline\hline
    & $\theta$ & $\psi$ & $\delta_4$ & $~\delta_5~$ & $\delta_6$ & $\tan\beta$   & $m_H^{}(=m_A = m_{H^\pm})$ [GeV] \\\hline
BP1 & $-0.438\pi$&$0.125\pi$&$-0.250\pi$&0&$-0.250\pi$&1.5&300 \\\hline
BP2 & $0.223\pi$ &$-0.446\pi$ & $-0.250\pi$&0&$-0.250\pi$ &2.5&300 \\\hline
BP3 & $-0.438\pi$&$0.125\pi$&$-0.250\pi$&0&$-0.250\pi$&3&500 \\\hline
BP4 & $-0.125 \pi$ & $0.0625\pi$ & $-0.271\pi$  & 0 & $-0.0637\pi$ & 3	&500\\\hline\hline
	\end{tabular}
                \caption{Input parameters for each benchmark point. }
                \label{tab:bps}
	\end{center}
%
	\begin{center}
		\begin{tabular}{cccccccc}
			\hline\hline
    & $|\Delta m_K|$ [MeV]   & $|\Delta m_D|$  [MeV]  & $|\Delta m_{B_d}|$  [MeV] & $|\Delta m_{B_s}|$  [MeV] & ${\cal B}(B \to X_s\gamma)$  & $|d_n|/e$ [cm]    \\\hline
BP1 & $3.17\times 10^{-12}$  & $\simeq 0$             & $1.47\times 10^{-10}$     & $1.22\times 10^{-10}$     & $3.06\times 10^{-4}$ & $4.24 \times 10^{-27}$ \\\hline
BP2 & $3.82\times 10^{-13}$  & $\simeq 0$             & $4.50\times 10^{-11}$     & $8.52\times 10^{-10}$     & $3.26\times 10^{-4}$ & $1.05\times 10^{-26}$ \\\hline
BP3 & $2.70 \times 10^{-12}$ & $\simeq 0$             & $1.26 \times 10^{-10}$    & $1.05\times 10^{-10}$     & $3.29\times 10^{-4}$ & $1.25\times 10^{-26}$\\\hline
BP4 & $1.87\times 10^{-13}$  & $2.82 \times 10^{-13}$ & $2.13 \times 10^{-10}$     & $3.06 \times 10^{-10}$    & $3.57\times 10^{-4}$ & $1.31 \times 10^{-26}$
	\\\hline\hline
	\end{tabular}
                \caption{Predictions of flavor observables for each benchmark point.}
                \label{tab:flavor}
	\end{center}
%
	\begin{center}
		\begin{tabular}{cccc|ccc|ccc}
			\hline\hline
    & \multicolumn{3}{c|}{${\cal B}(H)$ [\%]} & \multicolumn{3}{c|}{${\cal B}(A)$ [\%]} & \multicolumn{3}{c}{${\cal B}(H^\pm)$ [\%]}  \\\hline
BP1 & 98.3 ($tu$) & 0.8 ($gg$) & 0.6 ($tc$) & 98.8 ($tu$) & 0.6 ($tc$)& 0.3 ($gg$) & 93.3 ($tb$) & 6.6 ($td$)& 0.04 ($ts$)\\\hline
BP2 & 56.0 ($tc$) & 41.0 ($tu$) & 2.4 ($gg$) & 57.0 ($tc$) & 41.7 ($tu$)& 0.8 ($gg$) & 97.7 ($tb$) & 1.3 ($ts$) & 0.9 ($td$) \\\hline
BP3 & 92.0 ($tt$) & 7.6 ($tu$) & 0.3 ($gg$)  & 95.8 ($tt$) & 4.2 ($tu$) & 0.02 ($tc$) & 96.0 ($tb$) & 3.9 ($td$) & 0.02 ($ts$)\\\hline
BP4 & 80.0 ($tu$) & 17.7 ($tc$)	& 2.0 ($tt$) & 78.6 ($tu$) & 17.4 ($tc$) & 3.8 ($tt$) & 78.4 ($td$) & 17.4 ($ts$) & 4.2 ($tb$)\\\hline\hline
	\end{tabular}
                \caption{Predictions of the biggest three decay branching ratios of $H$, $A$ and $H^\pm$ for each benchmark point. The decay mode is written in the parentheses. 
For the branching ratios of flavor violating modes of $H$ and $A$, we take the sum of possible two final states, e.g., $H/A \to t\bar{c}$ and $H/A \to c\bar{t}$. }
                \label{tab:decay}
	\end{center}
\end{table}

We briefly discuss decays of the additional Higgs bosons. 
In the case with $s_{\beta-\alpha} = 1$ and $m_H = m_A = m_{H^\pm}$, the additional Higgs bosons can mainly decay into a fermion pair and sub-dominantly decay into two gluons, 
where the latter is induced at one-loop level for $H$ and $A$. 
The other loop induced decays $H/A \to \gamma\gamma/Z\gamma$ and $H^\pm \to W^\pm \gamma/W^\pm Z$ are negligible~\cite{Aiko:2020ksl}. 
The analytic expressions for the relevant decay rates are given in Appendix~\ref{sec:decay-rate}. 

In Table~\ref{tab:bps}, we give four benchmark points (BP1-BP4) 
all of which are allowed by the constraints from various flavor experiments as shown in Table~\ref{tab:flavor}. 
In Table~\ref{tab:decay}, we show the predictions of the branching ratios of $H$, $A$ and $H^\pm$ for each benchmark point. 
At the BP1 and BP2, the masses of $H$ and $A$ are taken to be 300 GeV, so that they cannot decay into a top-quark pair. 
In such a case, these additional Higgs bosons can typically decay into $b\bar{b}$ or $\tau^+\tau^-$ in the four types of the 2HDMs~\cite{Aoki:2009ha}, 
but they dominantly decay into $ts$ or $tc$ via the flavor violating interaction in our scenario. 
In these benchmark points, $H^\pm$ dominantly decay into $tb$ and sub-dominantly decay into $td$ and $ts$. 
At the BP3 and BP4, the masses of the additional Higgs bosons are taken to be 500 GeV, so that they can decay into a top-quark pair. 
We find that even in such a heavier case, $H$ and $A$ ($H^\pm$) mainly decay into $tu$ and $tc$ ($td$ and $ts$) at the BP4. 
Therefore, the flavor violating decays of the additional Higgs bosons can be important to test our scenario. 
Recently in Ref.~\cite{Hou:2020chc}, 
LHC phenomenology for flavor violating Higgs bosons has been discussed in the 2HDM without a $Z_2$ symmetry. 
It has been shown that the case with the magnitude of a parameter $\rho_{tc}$, corresponding to our $\sqrt{2}m_t(\Gamma_A^u)_{32}/v$, to be larger than about 0.5 
and with the mass range of $350 \lesssim m_{H/A} \lesssim 570$ GeV 
can be proved with 5$\sigma$ level from the three top process assuming 14 TeV for the collision energy and 3000 fb$^{-1}$ for the integrated luminosity~\cite{Hou:2020chc}. 
We note that $|\rho_{tc}|$ is given to be 0.03, 0.28, 0.044 and 0.55 at the BP1, BP2, BP3 and BP4, respectively. 
Thus, our scenario, especially BP4, can be proved at LHC.

\section{Conclusions \label{sec:conc}}

We have discussed a CPV 2HDM emerging from 3-3-1 models at the EW scale, which are motivated to explain the three generation structure of chiral fermions. 
All the non-SM fermions and gauge bosons are decoupled from the theory in our approach. 
We have shown that 3-3-1 models categorized as Class-I deduce the same structure of the 2HDM, in which
flavor violating Higgs-quark couplings appear, and they contain additional CPV phases. 
The Yukawa couplings for quarks can be parameterized not only by $\tan\beta$ but also the mixing parameters $\theta$, $\psi$ and the new phases $\delta_{4,5,6}$ which
arise from the bi-unitary transformation of up-type and down-type quarks. 
These new parameters do not appear in the softly-broken $Z_2$ symmetric 2HDMs, i.e., the Type-I, Type-II, Type-X and Type-Y 2HDMs, because of the flavor universal structure of the 
Yukawa interaction. 
If we take additional mixing and phase parameters to be zero, our scenario becomes Type-II 2HDM like as long as we neglect Yukawa interactions with the first and second generation fermions. 

We then have studied constraints on the parameter space from flavor observables such as the $B^0$-$\bar{B}^0$, $D^0$-$\bar{D}^0$, $K^0$-$\bar{K}^0$ mixings, the $B \to X_s \gamma$ decay and the EDMs particularly for the nEDM $d_n$. 
Because of the characteristic flavor structure, the neutral Higgs bosons do not contribute to the EDMs at one-loop and two-loop Barr-Zee diagrams. 
It has been clarified that the dominant contribution comes from charged Higgs boson loops at one-loop level. 
We have found that even in the case without new phases, i.e., $\delta_{4,5,6} = 0$
the charged Higgs boson contribution to $d_n$ can be sizable due to the CKM phase and the non-zero mixing angles $\theta$ and $\psi$, which can be of order $10^{-27}$ in the allowed parameter region. 
We also have found that the constraint from the $B \to X_s \gamma$ decay can significantly be relaxed as compared with the Type-II 2HDM in the case with $\theta,~\psi\neq 0$,
and have shown that a light charged Higgs boson with a few hundred GeV is possible. 
For the case with non-zero CPV phases, i.e., $\delta_{4,5,6} \neq 0$, we see that 
the effect of these phases on $d_n$ is constructive in a portion of the parameter space which are already excluded by the current measurement of $d_n$. 
On the other hand, there are regions where the new phases give a destructive effect, in which new phases of order one can be allowed under the flavor constraints. 
Such sizable CPV phases can be indirectly be tested at future EDM experiments such as the n2EDM experiment, and can directly be tested at collider experiments such as the high-luminosity LHC and future electron-positron colliders.

\begin{acknowledgments}
The work of KY was supported in part by Grant-in-Aid for Early-Career Scientists, No.~19K14714. 
\end{acknowledgments}

\begin{appendix}

\section{Masses of Higgs bosons \label{sec:app1}}

We discuss details of the mass spectrum for scalar bosons in the model with $\zeta = -1/\sqrt{3}$ whose particle content is given in Table~\ref{table1}. 

There are five neutral components in the three Higgs triplets, which can be parameterized as
\begin{align}
\phi_i^0 = \frac{1}{\sqrt{2}}(v_i + h_i + ia_i)~~(i=1,2,3),\quad 
\eta_j^0 = \frac{1}{\sqrt{2}}(v_j' + \eta_j^R  + i\eta_j^I)~~(j=2,3). 
\end{align}
We here introduce the VEVs for the $\eta_j^0$ fields, but either $v_2'$ and $v_3'$ can be taken to be zero by the field redefinition of $\Phi_2$ and $\Phi_3$ without loss of generality. 
We thus set $v_2' = 0$. 
From the tadpole condition for $\eta_2^R$, we obtain 
\begin{align}
m_{23}^2v_3 + \frac{v_3's_\beta}{2} \left(vv_3\rho_{23} + \frac{vM^2}{v_3}\right) = 0. 
\end{align}
For the case with $m_{23}^2 = 0$\footnote{This can be regarded by the choice of the soft-breaking term of the $U(1)'$ symmetry such that the subgroup $\tilde{Z}_2$ survives, where 
its charge is defined by $(-1)^{|Q'|/q}$~\cite{Das:2020pai}. }, this can be satisfied by taking $v_3' = 0$ for arbitrary values of $v$ and $v_3$. 
We take this configuration throughout the paper. 
In this setup, the tadpole condition for $\eta_3^R$ is automatically satisfied, while those for $h_{1,2,3}$ states are expressed under the assumption with $v,v_3 \neq 0$ as   
\begin{align} 
	&m_1^2 - M^2 s_{\beta }^2 + \frac{\lambda_{1}}{2}v^2c_{\beta }^2 + \frac{\lambda_{12}}{2} v^2  s_{\beta }^2+ \frac{\lambda_{13}}{2} v_3^2=0,\\
	&m_2^2 - M^2 c_{\beta }^2 + \frac{\lambda_{12}}{2} v^2  c_{\beta }^2+ \frac{\lambda_2}{2} v^2s_{\beta }^2+ \frac{\lambda_{23}}{2} v_3^2=0,\\
	&m_3^2 - M^2\frac{v^2}{v_3^2} c_{\beta }^2 s_{\beta }^2 + \frac{\lambda_{13}}{2} v^2  c_{\beta }^2 +  \frac{\lambda_{23}}{2} v^2 s_{\beta }^2 + \frac{\lambda_3}{2} v_3^2=0,
\end{align}
where $M^2 \equiv \mu v_3 /(\sqrt{2}c_\beta s_\beta)$. 

The mass eigenstates for the $Z_2^{\rm rem}$-even scalars can be defined as 
\begin{align}
\begin{pmatrix}
\phi_1^\pm \\
\phi_2^\pm \\
\end{pmatrix}
&=R(\beta)
\begin{pmatrix}
G^\pm \\
H^\pm \\
\end{pmatrix},~
R(x)=\begin{pmatrix}
\cos x& \sin x\\ 
-\sin x & \cos x
\end{pmatrix}. 
\\
\begin{pmatrix}
a_1 \\
a_2 \\
a_3
\end{pmatrix}
&=
\begin{pmatrix}
-\frac{v  c_{\beta }}{ \sqrt{v_3^2 + v^2 c_{\beta }^2}} & -c_\beta & \frac{v_3s_\beta}{\sqrt{v_3^2 + s_\beta^2 c_\beta^2 v^2}}\\
0 &  s_\beta & \frac{v_3c_\beta}{\sqrt{v_3^2 + s_\beta^2 c_\beta^2 v^2}}\\
\frac{v_3}{\sqrt{v_3^2 + v^2c_\beta^2}} & 0 & \frac{vs_\beta c_\beta}{\sqrt{v_3^2 + s_\beta^2 c_\beta^2 v^2}}
\end{pmatrix}
\begin{pmatrix}
G_1^0 \\
G_2^0 \\
A     \\
\end{pmatrix},~  
\begin{pmatrix}
h_1 \\
h_2 \\
h_3 
\end{pmatrix}
=R_H
\begin{pmatrix}
H \\
h \\
H_S 
\end{pmatrix}. \label{eq:mata} 
\end{align}
where $G^\pm$, $G_1^0$ and $G_2^0$ are the Nambu-Goldstone (NG) bosons, and 
$h$ can be identified with the discovered Higgs boson with a mass of 125 GeV. 
The matrix $R_H$ is the $3\times 3$ orthogonal matrix.
We note that the eigenvectors for $G_1^0$ and $G_2^0$, corresponding to the first and second column vectors of the $3\times 3$ matrix in Eq.~(\ref{eq:mata}), respectively,
are not orthogonal, so that the orthogonal NG states are obtained by taking an appropriate basis transformation of $(G_1^0,G_2^0)$. 
The masses of physical scalar states are calculated as 
\begin{align}
m_{H^\pm}^2 &= M^2 + \frac{\rho_{12}}{2}v^2, \quad m_A^2 = M^2 \left(1 + \frac{v^2}{v_3^2}s_\beta^2 c_\beta^2\right), \quad m_{H_i}^2 = (R_H^T M_H^2 R_H)_{ii}, 
\end{align} 
where $M_H^2$ is the squared mass matrix for CP-even Higgs bosons in the basis of $(h_1,h_2,h_3)$ given as 
\begin{align}
M_H^2 = 
\begin{pmatrix}
v^2c_\beta^2\lambda_1 + M^2 s_\beta^2 & (v^2 \lambda_{12} - M^2)s_\beta c_\beta & \frac{v}{v_3}c_\beta (v_3^2 \lambda_{13} - M^2 s_\beta^2) \\
& v^2s_\beta^2\lambda_2 + M^2 c_\beta^2 & \frac{v}{v_3}s_\beta (v_3^2 \lambda_{23} - M^2 c_\beta^2) \\
&& v_3^2 \lambda_3 + \frac{v^2}{v_3^2}M^2s_\beta^2 c_\beta^2
\end{pmatrix}. 
\end{align}
For the $Z_2^{\text{rem}}$-odd fields, the mass eigenstates are defined as follows:
\begin{align}
	&	\left(\begin{matrix}
		\eta^\pm_3\\
		\eta_1^\pm
	\end{matrix}\right)=
	R(\gamma)\left(\begin{matrix}
		G^\pm_\eta\\
		\eta^\pm
	\end{matrix}\right),~
\left(\begin{matrix}
		\eta_3^0 \\
		\eta_2^0
	\end{matrix}\right)=R(\delta)\left(\begin{matrix}
		G_\eta^0\\
		\eta^0
 	\end{matrix}\right)
\end{align}
where $G_\eta^\pm$ and $G_\eta^0$ are the NG bosons, $\tan \gamma \equiv v_1/v_3$ and $\tan \delta \equiv v_2/v_3$. 
The masses of these scalar bosons are given by
\begin{align}
m_{\eta^\pm}^2 = \left(1 + \frac{v^2}{v_3^2}c_\beta^2 \right)\left(M^2 s_\beta^2 + \frac{v_3^2}{2} \rho_{13}\right), \quad 
m_{\eta^{0}}^2 =\left(1 + \frac{v^2}{v_3^2}s_\beta^2 \right)\left(M^2 c_\beta^2 + \frac{v_3^2}{2} \rho_{23}\right). 
\end{align}

Let us take the large VEV limit, i.e., $v_3 \gg v$ with keeping the $M^2$ parameter to be the size of $v^2$. 
In this case, the $\eta^\pm$  and $\eta^0$ are decoupled as their masses contain the $v_3^3$ term. 
On the other hand, the matrix $M_H^2$ takes a block diagonal form with small corrections proportional to $v^2/v_3^2$ or $M^2/v_3^2$ as
\begin{align}
M_H^2 = 
\begin{pmatrix}
v^2c_\beta^2\lambda_1 + M^2 s_\beta^2 & (v^2 \lambda_{12} - M^2)s_\beta c_\beta & 0\\
& v^2s_\beta^2\lambda_2 + M^2 c_\beta^2 &                                      0 \\
0&0& v_3^2 \lambda_3 + \frac{v^2}{v_3^2}M^2s_\beta^2 c_\beta^2
\end{pmatrix} + {\cal O}\left(\frac{v^2}{v_3^2},\frac{M^2}{v_3^2}\right). 
\end{align}
The orthogonal matrix $R_H$ is then approximately expressed as 
\begin{align}
R_H=\begin{pmatrix}
\cos\alpha + {\cal O}(\epsilon^2) & -\sin\alpha + {\cal O}(\epsilon^2) &  {\cal O}(\epsilon)\\
\sin\alpha + {\cal O}(\epsilon^2) & \cos\alpha + {\cal O}(\epsilon^2) &  {\cal O}(\epsilon)\\
 {\cal O}(\epsilon) & {\cal O}(\epsilon) & 1 + {\cal O}(\epsilon^2)
\end{pmatrix}
\end{align}
with $\epsilon = v/v_3$. 
Neglecting the ${\cal O}(\epsilon)$ term, the masses of CP-even Higgs bosons are expressed as 
\begin{align}
  \label{44}
  m^2_H&=(M_H^{\prime 2})_{11} c_{\beta -\alpha }^2-2(M_H^{\prime 2})_{12} c_{\beta -\alpha } s_{\beta -\alpha }+(M_H^{\prime 2})_{22} s_{\beta -\alpha }^2, \\
  m^2_h&=(M_H^{\prime 2})_{11} c_{\beta -\alpha }^2+2(M_H^{\prime 2})_{12} c_{\beta -\alpha } s_{\beta -\alpha }+(M_H^{\prime 2})_{22} s_{\beta -\alpha }^2,  \\
  m^2_{H_S}&=v_3^2\lambda_3, 
\end{align}
and the mixing angle $\beta-\alpha$ is expressed as
\begin{align}
	\tan 2(\beta-\alpha) = \frac{-2(M_H^{\prime 2})_{12} }{(M_H^{\prime 2})_{11} -(M_H^{\prime 2})_{22} }. 
\end{align}
In the above expressions,  $M_H^{\prime 2}$ is the $2\times 2$ mass matrix in the basis of $R(-\beta)(h_1,h_2)^T$: 
\begin{align}
(M^{\prime 2}_H)_{11}&=v^2 \left(\lambda _1 c_{\beta }^4+\lambda _2 s_{\beta }^4 +2 \lambda_{12} c_{\beta }^2 s_{\beta }^2\right),\\
(M^{\prime 2}_H)_{12}&=v^2 c_{\beta } s_{\beta } \left(\lambda_2s^2_\beta  -\lambda_1 c^2_\beta +\lambda_{12} c_{2\beta}\right), \\
(M^{\prime 2}_H)_{22}&=M^2 +c_{\beta }^2 s_{\beta }^2 v^2\left(\lambda _1+\lambda _2 -2 \lambda_{12} \right). 
\end{align}
We see that at the large $v_3$ limit, the particle content of the Higgs sector coincides with 
that of the 2HDM, i.e., $H^\pm$, $A$, $H$ and $h$. 

\section{Other Class-I models \label{sec:app2}}

\begin{table}[t]
	\renewcommand\arraystretch{0.8}
	\begin{center}
		\begin{tabular}{ccccc}
			\hline
			Fields&~~$SU(3)_C\otimes SU(3)_L \otimes U(1)_X\otimes U(1)'$~~&~~$Z_2^{\rm rem}$~~&Components& \\
			\hline
			$Q^a_L$&$\left(\bm{3},\overline{\bm{3}},1/3,0\right)$& $(+,+,-)$ &$\left(d^a_L,-u^a_L,U^a_L\right)^T$&\\
			$Q^3_L$&$\left(\bm{3},\bm{3},0,0\right)$&$(+,+,-)$&$\left(t_L,b_L,D_L\right)^T$&\\
			$u^i_R$&$\left(\bm{3},\bm{1},+2/3,q\right)$&$+$&$u^i_R$&\\
			$d^i_R$&$\left(\bm{3},\bm{1},-1/3,-q\right)$&$+$&$d^i_R$&\\
			$U_R^a$&$\left(\bm{3},\bm{1},+2/3,-2q\right)$&$-$&$U_R^a$&\\
			$D_R$  &$\left(\bm{3},\bm{1},-1/3,2q\right)$&$-$&$D_R$&\\
			$L^i_L$&$\left(\bm{1},\bm{3},-2/3,0\right)$&$(+,+,-)$&$\left(\nu^i_L,e^i_L, E_L^i \right)^T$&\\
			$e^i_R$&$\left(\bm{1},\bm{1},-1,-q\right)$&$+$&$e^i_R$&\\
			$E^i_R$&$\left(\bm{1},\bm{1},-1,2q\right)$&$-$&$E^i_R$&\\\hline	
			$\Phi_1$&$\left(\bm{1},\bm{3},-2/3,-q\right)$&$(+,+,-)$&$\left(\phi_1^0,\phi_1^-,\eta_1^- \right)$&\\
                        $\Phi_2$&$\left(\bm{1},\bm{3},+1/3,q\right)$& $(+,+,-)$&$\left(\phi_2^+,\phi_2^0,\eta_2^0\right)$&\\
	                $\Phi_3$&$\left(\bm{1},\bm{3},+1/3,-2q\right)$&$(-,-,+)$&$\left(\eta_3^+,\eta_3^0,\phi_3^0\right)$&\\	\hline
		\end{tabular}
		\caption{Particle content of the model with $\zeta = +1/\sqrt{3}$, where $E^i$ are the extra lepton with the electric charge of $-1$. 
                  }
		\label{model2}
	\end{center}
\end{table}

\begin{table}[h]
	\renewcommand\arraystretch{0.8}
	\begin{center}
		\begin{tabular}{ccccc}
			\hline
			Fields&~~$SU(3)_c\otimes SU(3)_L \otimes U(1)_X$~~&~~$Z_2^{\rm rem}$~~&Components& \\
			\hline
			$Q^a_L$&$\left(\bm{3},\overline{\bm{3}},-1/3\right)$& $(+,+,-)$ &$\left(d^a_L,-u^a_L,J^a_L\right)^T$&\\
			$Q^3_L$&$\left(\bm{3},\bm{3},+2/3\right)$&$(+,+,-)$&$\left(t_L,b_L,K_L\right)^T$&\\
			$u^i_R$&$\left(\bm{3},\bm{1},+2/3\right)$&$+$&$u^i_R$&\\
			$d^i_R$&$\left(\bm{3},\bm{1},-1/3\right)$&$+$&$d^i_R$&\\
			$K_R$&$\left(\bm{3},\bm{1},+5/3\right)$&$-$&$K_R$&\\
			$J^a_R$&$\left(\bm{3},\bm{1},-4/3\right)$&$-$&$J^a_R$&\\
			$L^i_L$&$\left(\bm{1},\bm{3},0\right)$&$(+,+,-)$&$\left(\nu^i_L,e^i_L,(e_R^i)^c\right)^T$&\\\hline	
                        $\Phi_1$&$\left(\bm{1},\bm{3},0\right)$& $(+,+,-)$&$\left(\phi_1^0,\phi_1^-,\eta_1^+\right)$&\\
			$\Phi_2$&$\left(\bm{1},\bm{3},1\right)$&$(+,+,-)$&$\left(\phi_2^+,\phi_2^0,\eta_2^{++} \right)$&\\
	                $\Phi_3$&$\left(\bm{1},\bm{3},-1\right)$&$(-,-,+)$&$\left(\eta_3^-,\eta_3^{--},\phi_3^0\right)$&\\	\hline
		\end{tabular}
		\caption{Particle content for the model with $\zeta = -\sqrt{3}$, where $J^a$ and $K$ are exotic quarks with the electric charge of $-4/3$ and $+5/3$, respectively. } 
		\label{model3}
	\end{center}
\end{table}

\begin{table}[h]
	\renewcommand\arraystretch{0.8}
	\begin{center}
		\begin{tabular}{ccccc}
			\hline
			Fields&~~$SU(3)_c\otimes SU(3)_L \otimes U(1)_X$~~&~~$Z_2^{\rm rem}$~~&Components& \\
			\hline
			$Q^a_L$&$\left(\bm{3},\overline{\bm{3}},2/3\right)$& $(+,+,-)$ &$\left(d^a_L,-u^a_L,K^a_L\right)^T$&\\
			$Q^3_L$&$\left(\bm{3},\bm{3},-1/3\right)$&$(+,+,-)$&$\left(t_L,b_L,J_L\right)^T$&\\
			$u^i_R$&$\left(\bm{3},\bm{1},+2/3\right)$&$+$&$u^i_R$&\\
			$d^i_R$&$\left(\bm{3},\bm{1},-1/3\right)$&$+$&$d^i_R$&\\
			$K_R$&$\left(\bm{3},\bm{1},+5/3\right)$&$-$&$K_R^a$&\\
			$J^a_R$&$\left(\bm{3},\bm{1},-4/3\right)$&$-$&$J_R$&\\
			$L^i_L$&$\left(\bm{1},\bm{3},-1\right)$&$(+,+,-)$&$\left(\nu^i_L,e^i_L,F_L^i \right)^T$&\\
			$e^i_R$&$\left(\bm{1},\bm{1},-1\right)$&$+$&$e^i_R$&\\
			$F^i_R$&$\left(\bm{1},\bm{1},-2\right)$&$-$&$F^i_R$&\\\hline	
                        $\Phi_1$&$\left(\bm{1},\bm{3},-1\right)$& $(+,+,-)$&$\left(\phi_1^0,\phi_1^-,\eta_1^{--}\right)$&\\
			$\Phi_2$&$\left(\bm{1},\bm{3},0\right)$&$(+,+,-)$&$\left(\phi_2^+,\phi_2^0,\eta_2^{-} \right)$&\\
	                $\Phi_3$&$\left(\bm{1},\bm{3},1\right)$&$(-,-,+)$&$\left(\eta_3^{++},\eta_3^{+},\phi_3^0\right)$&\\	\hline
		\end{tabular}
		\caption{Particle content for the model with $\zeta = +\sqrt{3}$, where $J$ and $K$ ($F$) are exotic quarks (leptons) 
with the electric charge of $-4/3$ and $+5/3$ ($-2$), respectively. } 
		\label{model4}
	\end{center}
\end{table}

The particle content of the other Class-I models with $\zeta = +1/\sqrt{3}$, $-\sqrt{3}$ and $+\sqrt{3}$ are shown in Tables~\ref{model2}, 
\ref{model3} and \ref{model4}, respectively. 
The indices $i$ and $a$ run over 1-3 and 1-2, respectively. 
For the models with $\zeta = \pm\sqrt{3}$, we do not need to impose the $U(1)'$ symmetry, because all the extra fermions have different electric charges from that of SM fermions, and thus
there is no mixing among SM and extra fermions. 
In the model with $\zeta = -\sqrt{3}$, the remnant $Z_2^{\rm rem}$ symmetry does not appear, because the SM right-handed charged leptons are embedded into the third component of the lepton triplet. 
In other words, the anti-symmetric Yukawa interaction $\overline{L_L^c}L_L\Phi_1$ breaks the remnant symmetry, by which $\eta$ particles can decay into SM particles. 
On the other hand, in the model with $\zeta = +\sqrt{3}$, the remnant $Z_2^{\rm rem}$ symmetry appears, and its charges can be extracted from interaction terms of component fields, which are shown in Table~\ref{model4}. 
However, it has to be explicitly broken, because there is no neutral $Z_2^{\rm rem}$-odd particle, i.e., it gives rise to a stable charged particle which has been excluded by cosmological observations. 
A simple way to break the $Z_2^{\rm rem}$ symmetry is to introduce doubly-charged singlet scalars $S^{++} \sim ({\bm 1},{\bm 1},+2)$, by which 
$\Phi_3^\dagger\Phi_1 S^{++}$ and $\overline{e_R^c}e_R S^{++}$ terms can be constructed, and these allow $Z_2^{\rm rem}$-odd particles to decay into SM particles.

\section{Decay rates of the Higgs bosons \label{sec:decay-rate}}

The decays rates into a pair of fermions are given by 
\begin{align}
\begin{split}
\Gamma(H \to \bar{q}_iq_j) & =\frac{N_c}{8\pi}\frac{m_H^3}{v^2}\left|(\Gamma_H^q)_{ij}\right|^2 \left[\frac{r_{H}^{q_i} + r_{H}^{q_j}}{2}(1-r_{H}^{q_i} - r_{H}^{q_j}) 
- 2r_{H}^{q_i}r_{H}^{q_j}\right]\sqrt{F_{\text{2-body}}(r_{H}^{q_i},r_{H}^{q_j})}, \\
\Gamma(H \to \bar{\ell}\ell) & =\frac{1}{8\pi}\frac{m_H^3}{v^2}r_H^\ell(c_{\beta-\alpha} + \tan\beta s_{\beta-\alpha})^2 \sqrt{F_{\text{2-body}}(r_H^\ell,r_H^\ell)}^3, \\
\Gamma(A \to \bar{q}_iq_j) & =\frac{N_c}{8\pi}\frac{m_A^3}{v^2}\left|(\Gamma_A^q)_{ij}\right|^2 \left[\frac{r_A^{q_i} + r_A^{q_j}}{2}(1 - r_A^{q_i} - r_A^{q_j}) + 2r_A^{q_i}r_A^{q_j}\right]\sqrt{F_{\text{2-body}}(r_A^{q_i},r_A^{q_j})}, \\
\Gamma(A \to \bar{\ell}\ell) & =\frac{1}{8\pi}\frac{m_A^3}{v^2}r_A^\ell(c_{\beta-\alpha} + \tan\beta s_{\beta-\alpha})^2 \sqrt{F_{\text{2-body}}(r_A^\ell,r_A^\ell)}, \\
\Gamma(H^+ \to \bar{d}_iu_j) & =\frac{N_c}{8\pi}\frac{m_{H^\pm}^3}{v^2}\Bigg\{[|(\Gamma_A^dV_{\rm CKM}^\dagger)_{ij}|^2r_{H^\pm}^{d_i} + |(V_{\rm CKM}^\dagger \Gamma_A^u)_{ij}|^2r_{H^\pm}^{u_j} ](1 - r_{H^\pm}^{d_i} - r_{H^\pm}^{u_j}) \\
& + 4r_{H^\pm}^{d_i}r_{H^\pm}^{u_j} \text{Re}[(\Gamma_A^dV_{\rm CKM}^\dagger)_{ij}^*(V_{\rm CKM}^\dagger \Gamma_A^u)_{ij}]\Bigg\}\sqrt{F_{\text{2-body}}(r_{H^\pm}^{d_i},r_{H^\pm}^{u_j})}, \\
\Gamma(H^+ \to \bar{\ell}\nu) & =\frac{1}{8\pi}\frac{m_{H^\pm}^3}{v^2}r_{H^\pm}^\ell \tan^2\beta\sqrt{F_{\text{2-body}}(r_{H^\pm}^\ell,0)} ,
\end{split} \label{eq:decay}
\end{align}
where $r_X^Y = m_Y^2/m_X^2$ and $N_c = 3$. 
In Eq.~(\ref{eq:decay}), $F_{\text{2-body}}$ is the 2-body phase space function given as 
\begin{align}
F_{\text{2-body}}(x,y) = [(1-x-y)^2 -4xy]\Theta(1 - \sqrt{x} - \sqrt{y}),  
\end{align}
with $\Theta$ being the Heaviside step function. 
The neutral Higgs bosons $H$ and $A$ can also decay into two gluons at one-loop level, and its decay rate is given by 
\begin{align}
\begin{split}
\Gamma(H \to gg) & = \frac{\alpha_s^2}{2\pi^3}\frac{m_H^3}{v^2}\left|r_H^t(\Gamma_H^u)_{33}I_t +r_H^b(\Gamma_H^d)_{33} I_b \right|^2, \\
\Gamma(A \to gg) & = \frac{\alpha_s^2}{2\pi^3}\frac{m_A^3}{v^2}\left|r_A^t(\Gamma_A^u)_{33}J_t +r_A^b(\Gamma_A^d)_{33} J_b \right|^2,
\end{split} \label{eq:decay2}
\end{align}
where 
\begin{align}
I_q &= 1 + (1 - 4r_H^q)f(r_q), \quad J_q = r_A^qf(r_q), 
\end{align}
with
\begin{align}
f(x) &= \arcsin^2\left[\frac{1}{2\sqrt{x}}\right],~\text{for}~x\geq \frac{1}{4}, \\
f(x) &= -\frac{1}{4}\left[\ln \left(\frac{1+\sqrt{1-4x}}{1-\sqrt{1-4x}}\right)    -i\pi\right]^2,~\text{for}~x < \frac{1}{4}. 
\end{align}

\end{appendix}

\bibliography{ref}
\end{document}